\documentclass[11pt]{article}
\pdfoutput=1

\usepackage{jheppub}
\usepackage{url,comment}
\usepackage[utf8x]{inputenc}
\usepackage{graphicx}
\usepackage{mathtools}

\usepackage{enumitem}
\usepackage{slashed}
\usepackage{url}
\usepackage[T1]{fontenc}
\usepackage{bm}
\usepackage{color}

\newcommand{\beq}{\begin{equation}}
\newcommand{\eeq}{\end{equation}}
\newcommand{\beqa}{\begin{eqnarray}}
\newcommand{\eeqa}{\end{eqnarray}}
\newcommand{\lp}{\left(}
\newcommand{\rp}{\right)}

\newcommand{\veva}[1]{\left< #1\right>}

\newcommand{\order}[1]{\mathcal{O}(#1)}
\DeclareMathOperator{\Tr}{Tr}

\newcommand{\keV}{\rm keV}
\newcommand{\MeV}{\rm MeV}

\newcommand{\TeV}{\rm TeV}

\newcommand{\Cincy}{Department of Physics, University of Cincinnati, Cincinnati, OH, USA}
\newcommand{\SantaCruz}{Santa Cruz Institute for Particle Physics, University of California, Santa Cruz, CA 95064, USA}

\title{Light Dark Matter from Entropy Dilution}
\author[a]{Jared A.~Evans,}
\author[b]{Akshay Ghalsasi,}
\author[b]{Stefania Gori,}
\author[a]{Michele Tammaro,}
\author[a]{Jure Zupan}
\affiliation[a]\Cincy
\affiliation[b]\SantaCruz

\emailAdd{jaredaevans@gmail.com}
\emailAdd{akshay.ghalsasi@gmail.com}
\emailAdd{sgori@ucsc.edu}
\emailAdd{tammarme@mail.uc.edu}
\emailAdd{zupanje@ucmail.uc.edu}

\begin{abstract}
{ 
We show that a thermal relic which decouples from the standard model (SM) plasma while relativistic can be a viable dark matter (DM) 
candidate, if the decoupling is followed by a period of entropy dilution that heats up the SM, but not the dark sector.
Such diluted hot relics can be as light as few keV, while accounting for the entirety of the DM, and not 
conflicting with cosmological and astrophysical measurements.  The requisite dilution can be achieved via decays of a  heavy state that dominates the energy budget of the universe in the early matter dominated era.
 The heavy state decays into the SM 
 particles, heats up the SM plasma, and dilutes the hidden sector.  The interaction required to equilibrate the two sectors in the early 
 universe places a bound on the maximum possible dilution as a function of the decoupling temperature.   As an example of diluted hot 
 relic DM we consider a light Dirac fermion with a heavy dark photon mediator. We present constraints on the model from terrestrial 
 experiments (current and future), astrophysics, and cosmology.
}
\end{abstract}

\begin{document}

\arxivnumber{nnnn.nnnnn}

\maketitle
\section{Introduction}

 There is overwhelming evidence that dark matter (DM) exists and makes up roughly a quarter of the universe's energy budget, based on its gravitational influence on myriad astrophysical and cosmological observables \cite{Buckley:2017ijx,Bertone:2004pz}.  Far less is known about non-gravitational interactions of DM.  Fairly feeble interactions are sufficient to bring DM into thermal equilibrium with the standard model (SM), so that some mechanism, typically annihilations,  must be introduced to reduce the DM abundance to the measured level.  Such interactions have yet to be observed, with increasingly stringent limits being imposed by a number of experiments looking for DM in direct detection, through indirect detection, and at colliders, as well as probes from astrophysics and cosmology. 

The constraints are especially severe in the case of light DM.   Thermal relic DM with a mass below an MeV is essentially ruled out \cite{Serpico:2004nm,Ho:2012ug,Boehm:2013jpa,Nollett:2013pwa,Nollett:2014lwa,Green:2017ybv,Depta:2019lbe,Sabti:2019mhn}, although a few exceptions do exist \cite{Green:2017ybv,Berlin:2017ftj,Berlin:2018ztp,Berlin:2019pbq}.  These constraints are almost completely relaxed for DM that is a \emph{diluted hot relic}.  A hot relic is in thermal equilibrium with the SM in the early universe, but decouples from the plasma, i.e., freezes out, while still relativistic.  This can be achieved for light DM that connects with a sizable coupling to the SM through a much heavier mediator particle, $m_{\rm med}\gg m_{\rm DM}$.  After the temperature of the universe drops below the mediator mass the interaction rate falls much faster than Hubble as the universe cools. As a result, light DM decouples while relativistic.    Normally, such a hot relic is subject to stringent constraints from cosmology. However, as we will show, the constraints can be relaxed if the hidden sector (HS) undergoes dilution during the cosmological evolution.

The dilution can be caused by the decay of a heavy state, e.g., a long-lived moduli, that dominates the energy budget of the universe during the relevant cosmological period.  By assumption, the heavy state decays predominantly into SM particles, heats up the SM plasma, and leaves the HS comparatively cold. In this mechanism, the  HS is sufficiently coupled to the SM in order to equilibrate in the early universe, but due to the mass of the heavy mediator decouples from the SM at later times, so that the entropy injected into the SM does not feed back into the HS. 

While a diluted hot relic can be as light as $m_{\rm DM}\sim$~ 4.4 keV, it could also be heavier than an MeV. 
On the contrary, typical thermal DM models with $s-$wave annihilation, for which the annihilation cross-section $\veva{\sigma v}$ is independent of velocity, have the lightest permissible mass of DM  constrained to be well above a GeV  by precision observations of the cosmic microwave background (CMB) power spectrum \cite{Slatyer:2015jla}.\footnote{Although models with coannihilation, coscattering, or forbidden annihilations can relax these constraints \cite{Griest:1990kh, DAgnolo:2015ujb, DAgnolo:2017dbv,DAgnolo:2018wcn}.}   While models with velocity suppressed annihilation cross-sections can evade the stringent CMB constraints, e.g., models with $p$-wave annihilation,  often these are subject to other substantial constraints.  For example, the case of direct freeze-out through the Higgs portal faces a number of additional stringent constraints from colliders, rare meson decays, and direct detection limits that together essentially rule out the model for DM lighter than the Higgs, $m_{\rm DM}<m_h$~\cite{Krnjaic:2015mbs}. 
In secluded annihilation or ``WIMP-next-door'' models \cite{Pospelov:2007mp}, on the other hand, the DM freezes out into light mediators that later decay into the SM. In this case $p$-wave annihilation easily allows for sub-GeV DM \cite{Evans:2017kti}. Since in the WIMP-next-door models the DM relic abundance is set entirely by the size of DM coupling to the mediators, the correct relic abundance is obtained even for very small couplings to the SM, and the dark sector is insulated from most of the experimental constraints \cite{Evans:2017kti}. The essential ingredient in all the models of this type is that the mediator is lighter than the DM. The main topic of the present paper is the opposite limit, light DM with a  heavy mediator, which is possible if DM is a diluted hot relic.  

\begin{figure}
  \centering
  \includegraphics[width=.49\linewidth]{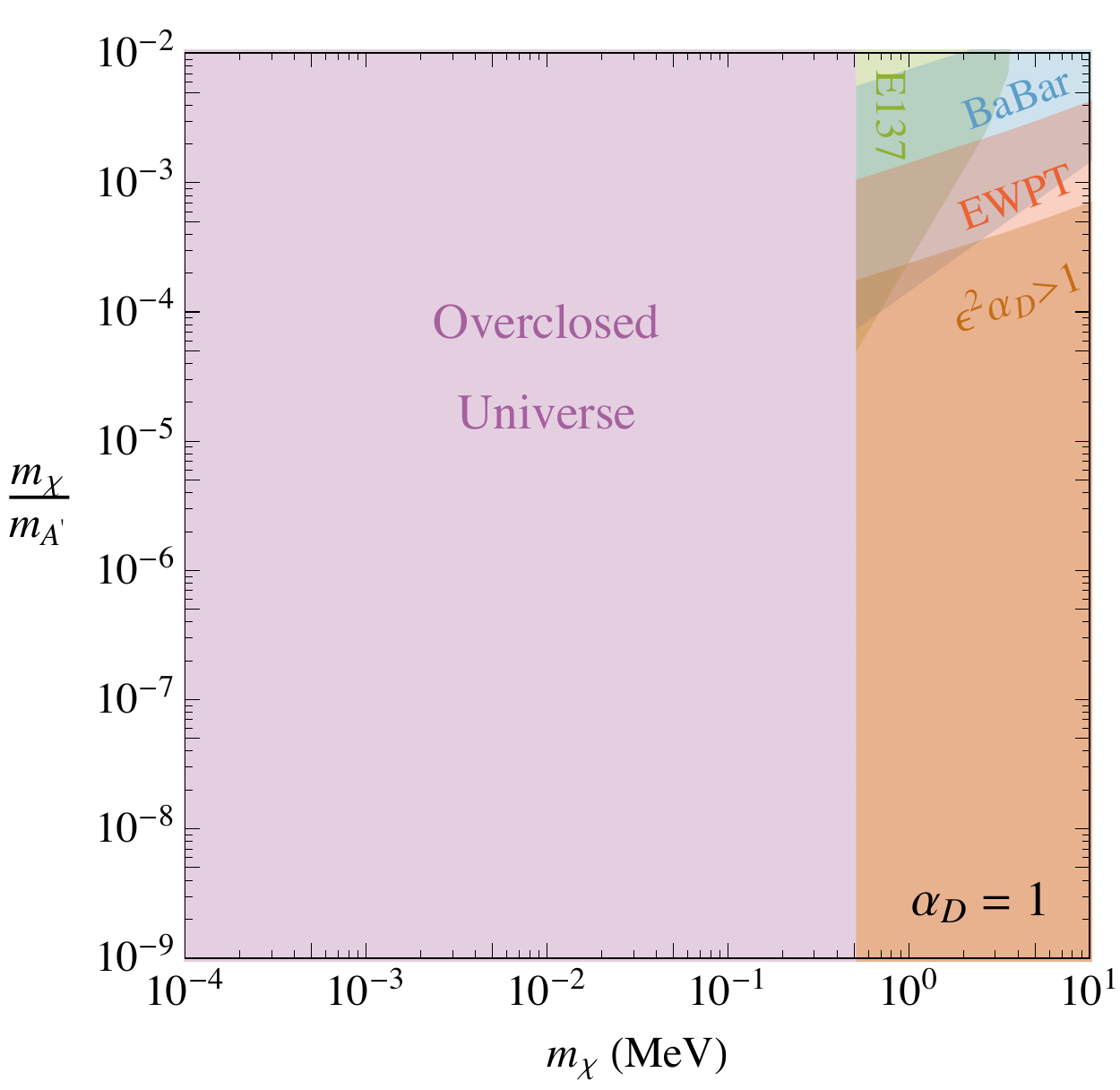}
  \includegraphics[width=.49\linewidth]{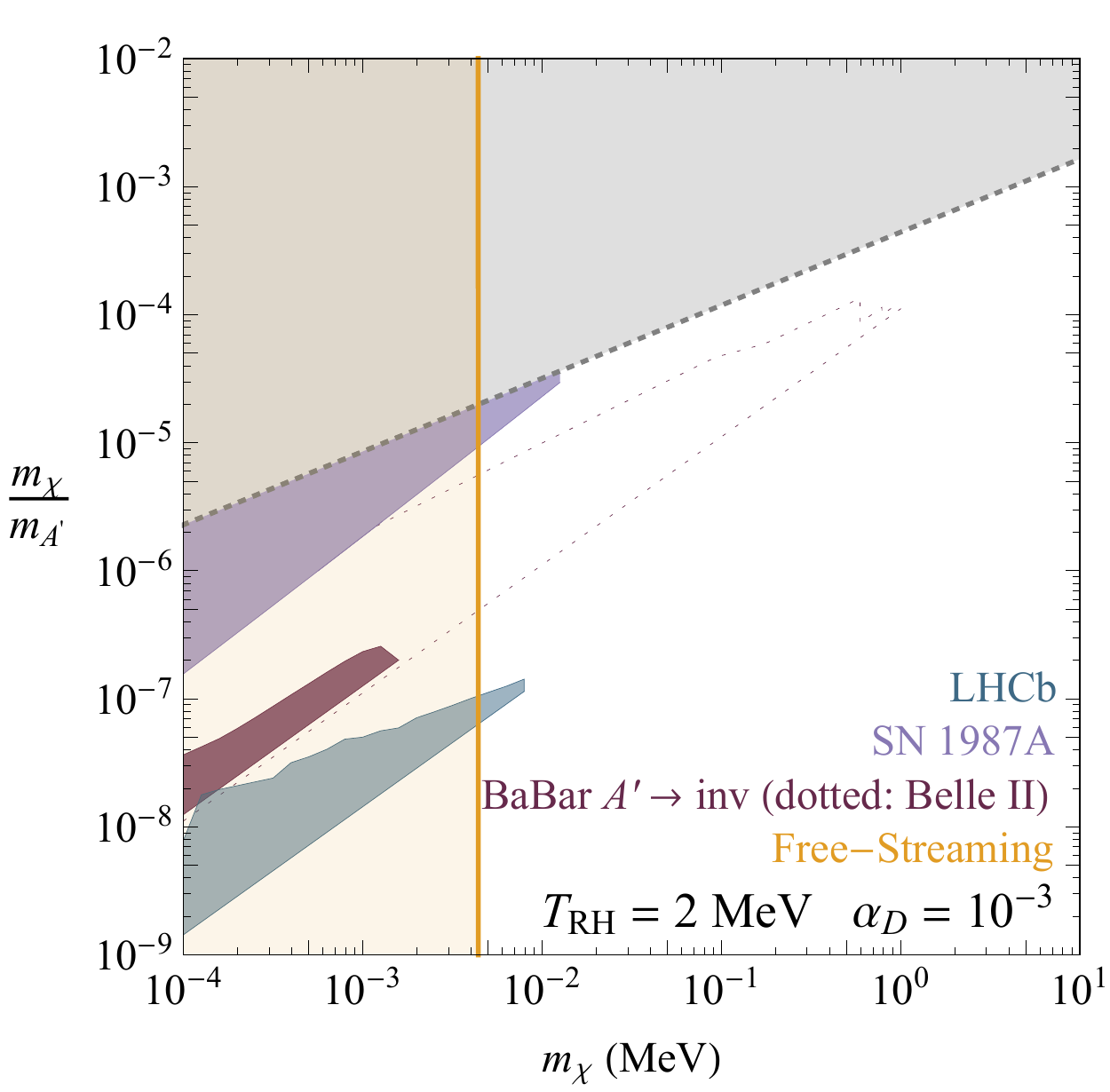}
\caption{Left panel shows the constraints for thermal freeze-out DM with masses between $100$~eV and $10~\MeV$, assuming no dilution. The constraints are significantly relaxed if the thermal history of the hidden sector contains a period of entropy dilution (right panel shows the constraints for the case of maximal dilution).  Above the dashed gray line, $T_D > m_{A'}/3$, and determinations of couplings are unreliable due to the resonance.  We do not extend into that region in this study. In both cases, DM is a Dirac fermion, with  the kinetically-mixed dark photon acting as the mediator.
}
\label{fig:pspace}
\end{figure}

Fig.~\ref{fig:pspace} shows the striking difference in the viable parameter space for two light DM scenarios with a  heavy mediator, a diluted hot relic DM (right panel) compared to the direct freeze-out without dilution (left panel). In both cases, DM is a Dirac fermion, with  the kinetically-mixed dark photon acting as the mediator. While for $m_\chi \lesssim m_e$ the thermal freeze-out without dilution leads to an overclosed universe, this is no longer the case for diluted hot relic DM. Most notably, the right panel of Fig.~\ref{fig:pspace} illustrates that  fermion DM with a ${\mathcal O}({\rm keV})$ mass does not need to be a sterile neutrino, and may well be a stable $Z_2$-odd diluted dark particle. 

For diluted hot relic DM, the lower bound on DM mass, $m_\chi\gtrsim5$ keV, is set by free-streaming constraints from the Lyman-$\alpha$ forest, which are somewhat stronger than the astrophysical Tremaine-Gunn constraints (see Section \ref{sec:HS:constr} for details). Since diluted hot relic DM models allow larger hierarchies between $m_{\rm DM}$ and $m_{\rm med}$, i.e., allow for heavier mediators, constraints from direct searches are typically relaxed. The remaining constraints are due to the observation of the neutrino pulse from SN1987A, searches for invisibly decaying dark photons at Babar and Belle II, and searches for promptly decaying dark photons at LHCb.  While in Sec. \ref{sec:vector} of this paper we focus on the Dirac fermion DM with a kinetically-mixed dark photon as a working example, the dilution mechanism we introduce is more general and can be applied to many other DM models, opening up the related parameter space. 

Thermal and non-thermal histories with dilution have been considered before in the literature
 in order to achieve the correct relic abundance.  These include models in which DM has smaller annihilation cross-section than the standard weakly-interacting-massive-particle (WIMP) \cite{Pallis:2004yy,Gelmini:2006pw,Gelmini:2006pq,Gelmini:2008sh,Arcadi:2011ev,Hamdan:2017psw},  ultra-heavy DM \cite{Bramante:2017obj,Cirelli:2018iax,Allahverdi:2018aux}, completely decoupled dark sectors \cite{Acharya:2009zt,Randall:2015xza}, sterile neutrinos \cite{Gelmini:2008fq,Bezrukov:2009th,Nemevsek:2012cd,Patwardhan:2015kga}, and axions \cite{Visinelli:2009kt,Ramberg:2019dgi}.  
 Our case contrasts with these works that involve thermalization followed by dilution in that we take into account the potentially important late-time interactions.  
 The dilution mechanism can be CP and baryon number violating, in which case it can also produce the baryon asymmetry \cite{Kohri:2009ka,Kane:2019nes}. 
 Alternatively, 
the baryon asymmetry could be generated early with a much larger asymmetry, while the observed value is obtained through dilution.  In this work, we are agnostic about the origin of the baryon asymmetry. 

The paper is organized as follows. In Section \ref{sec:dilution} we give a model independent overview of the dilution mechanism, followed by model independent cosmological and astrophysical constraints on diluted hot relic DM in Section \ref{sec:HS:constr}. In Section \ref{sec:vector} we apply these results to the specific example of a heavy vector portal that connects the HS to the SM. We also discuss the current and future terrestrial and astrophysical constraints on this model. Section \ref{sec:conclusion} contains our conclusions. Appendix \ref{app:SNE} contains the details about DM production in supernovae, and Appendix \ref{app:IT} the details on internal thermalization of the HS. 

\section{The Dilution Mechanism}
\label{sec:dilution}

One of the main goals of this paper is to derive the phenomenological consequences of possibly the simplest mechanism to dilute the hidden sector --  the injection of entropy from a late-decaying state.\footnote{Other dilution mechanisms are possible. For instance, dilution could be accomplished by a second era of inflation as in Ref. \cite{Lyth:1995ka,Davoudiasl:2015vba}.} For successful dilution, the late-decaying state should have the following properties:
 \begin{enumerate}
 \item it red-shifts in the same way that matter does,
 \item it dominates the universe's energy budget at high temperatures,
 \item it decays almost entirely into the SM states.
 \end{enumerate}
Examples include  a long-lived moduli \cite{Moroi:1999zb},\footnote{In general, string moduli behave differently than the field driving the dilution mechanism. In moduli decays, it is challenging to avoid sizable branching ratios into all sectors \cite{Moroi:1999zb}.}  late-decaying supersymmetric condensate \cite{Thomas:1995ze}, gravitino \cite{Moroi:1994rs},  inflaton \cite{Allahverdi:2002nb}, curvaton \cite{Moroi:2002rd}, dilaton \cite{Lahanas:2011tk}, $Q$-balls \cite{Fujii:2002kr}, or some other very heavy late-decaying thermal relic. The specific nature of the late-decaying state is  not very important for the cosmological evolution, as long as it satisfies the above three properties.  For ease of discussion, we will refer 
to the state that sources the dilution simply as the moduli.  

The salient features of the cosmological history can be distilled into five relevant parameters. Two parameters are related to the moduli itself: the decay rate of the moduli, $\Gamma_m$, and the co-moving energy stored in the moduli, $\Phi_m \equiv \rho_{\rm mod} a^3$, where $a$ is the scale factor. Three parameters are related to the hidden sector: the mass of the dark matter, $m_{\chi}$, the temperature, $T_D$, below which the SM and hidden sector have decoupled, and, lastly, $\tilde g_*$, the effective number of relativistic degrees of  freedom in the hidden sector at $T_D$.  From these parameters
\beq
\Gamma_m,~\Phi_m,~m_{\chi},~T_D,~\tilde g_*,
\eeq
a remarkable number of concrete predictions can be made. 
 
\subsection{A matter-dominated universe}
\label{sec:MDU}

In this subsection, we briefly review the history of the early universe  in the presence of an early matter dominated (MD) era and discuss the different periods. To simplify the discussion we will assume in this subsection that the plasma is entirely composed from the SM particles and ignore the HS plasma. In the next subsection we will then specialize to our case of a plasma that also contains the HS.  

The energy of the universe can be separated into two contributions that are constant during purely adiabatic expansion; the contribution $\Phi_m \equiv \rho_{\rm mod} a^3$ from the moduli, i.e.~matter that will eventually decay, and the contribution from radiation, $\Phi_R\equiv \rho_R a^4$. The Hubble expansion rate is therefore
\beq
H = \frac{\dot a}{a} = \frac{1}{M_{\rm pl}} \sqrt{\frac{8 \pi}{3} \lp \frac{\Phi_R}{a^4} +\frac{\Phi_{m}}{a^3} \rp},
\label{eq:Hub}
\eeq
where $M_{\rm pl}$ is the Planck mass. 
The universe evolves according to the Boltzmann equations 
\beqa
\dot\Phi_m &=& - \Gamma_{m} \Phi_m, \label{eq:diffeqbegin}\\
\dot \Phi_R &=& a \Gamma_{m} \Phi_m. \label{eq:diffeqend}
\eeqa

Given how these densities scale, 
at some early time, $t_{\rm eq}$, the matter and radiation energy densities would have been equal, i.e.,  $\rho_{{\rm mod},i}=\rho_{R,i}$, with $\rho_{{\rm mod},i}\equiv \rho_{\rm mod}(t_{\rm eq})$, $\rho_{R,i}\equiv \rho_{R}(t_{\rm eq})$. We can define the scale factor at $t_{\rm eq}$ to be $a_{\rm eq}\equiv 1$, so that $\Phi_{m,i} = \Phi_{R,i}$.  
The MD evolution then divides the cosmological history of the early universe into four characteristic epochs \cite{Co:2015pka},
\begin{equation*}
\begin{tabular}{rcl}
$T> T_{\rm eq}$ & & : early radiation domination (ERD), \\
$T_{\rm eq}>T> T_{\rm NA}$ & & : adiabatic matter domination (MD$_{\rm A}$),\\
$T_{\rm NA}> T>T_{\rm RH}$ & & : non-adiabatic matter domination  (MD$_{\rm NA}$), \\
$ T_{\rm RH} >T$ & & : radiation domination (RD). \\ 
\end{tabular}
\end{equation*}
Here $T_{\rm eq}$  is the temperature at matter-radiation equality, $T_{\rm NA}$ the temperature at which the non-adiabatic evolution of the SM plasma starts, and $T_{\rm RH}$ the reheat temperature after the decay of the moduli.  The evolution through the different epochs  in terms of scale factor $a$ are illustrated in Fig.~\ref{fig:schematicplot}.  However, note that the universe never has to have attained {$T_{\rm eq}$ in order for the dilution mechanism to function. This assumption should be viewed as a means to simplify the presentation.   The important aspect is the existence of an era of matter domination, while $t_{\rm eq}$ merely conveniently sets the clock for our discussion.

\begin{figure}[t]
\centering
 \includegraphics[scale=1.1]{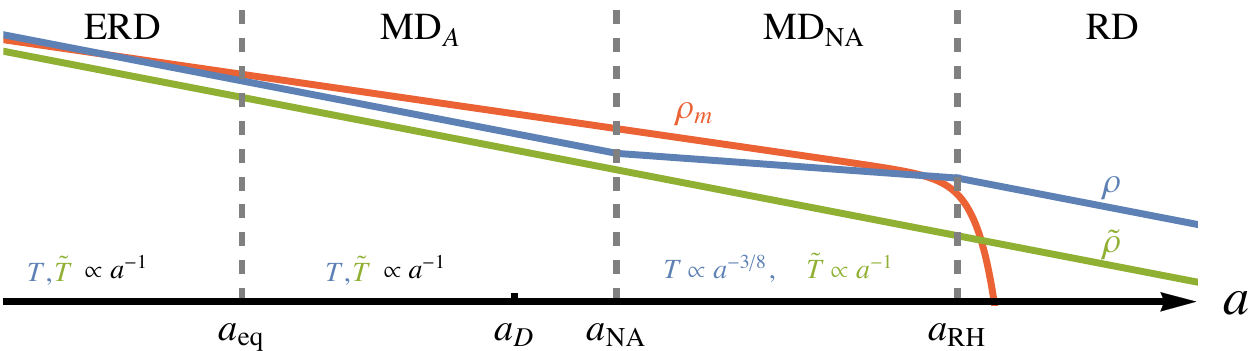}
\caption{The evolutions of the energy densities of the moduli, $\rho_m$ (red), the SM sector, $\rho$ (blue), and hidden sector, $\tilde \rho$ (green). The different periods are the early radiation domination phase (ERD), matter dominated adiabatic evolution (MD$_\text{A}$), matter dominated non-adiabatic evolution (MD$_\text{NA}$), and late radiation domination (RD), delineated by the scale factors $a_{X}$ corresponding to temperatures $T_{X}$.  At early matter--radiation equality, $a_{\rm eq}\equiv 1$, we have $\rho_m=\rho+\tilde\rho$.  At $a_{\rm NA}$, the energy density injected into the SM by the decaying moduli begins to exceed the red-shifted existing energy density.  At $a_{\rm RH}$, the moduli has mostly decayed and the universe is again radiation dominated.  At $a_D$, the SM and hidden sector plasmas decouple. In the plot this was taken to occur before $a_{\rm NA}$, see Section \ref{sec:HSdil} for the discussion of both cases, before and after  $a_{\rm NA}$.  The effects of energy leaking into $\tilde\rho$ from the hotter $\rho$ is not shown.  A schematic illustrating these effects is shown in Fig.~\ref{fig:schematicplotApp} of Appendix~\ref{app:IT}.
  \label{fig:schematicplot}}  
\end{figure}

The temperature $T_{\rm eq}$ of the SM plasma at the time of matter-radiation equality, $t_{\rm eq}$, when $\rho_{{\rm mod},i}=\rho_{R,i}$, is given by
 \beq
 T_{\rm eq} = \lp \frac{30 \, \Phi_m}{ \pi^2 g_{*}(T_{\rm eq})} \rp^{\frac 14}  \approx 1.32 \lp \frac{\Phi_m}{g_{*}(T_{\rm eq})}\rp^{\frac 14},
 \label{eq:Teq}
 \eeq
 where $g_{*}(T)$ is the effective number of relativistic degrees of freedom (d.o.f.)\ in the SM at temperature $T$.
 
 To estimate $T_{\rm NA}$, we rely on the fact that this epoch begins long before the bulk of the moduli decay, $t_{\rm NA} \ll \Gamma_m^{-1}$. During this period we can approximately take $\Phi_m\simeq \Phi_{m,i}=$const, so that Eqs.~(\ref{eq:Hub}~--~\ref{eq:diffeqend}) can be written as a single differential equation \cite{Evans:2016zau},
 \beq
 \label{eq:dPhiRda}
 \frac{d\Phi_R}{da} = \sqrt{\frac{3}{8\pi}} M_{pl} \Gamma_m  \frac{\Phi_m a^2}{\sqrt{a \Phi_m + \Phi_R}}\simeq \sqrt{\frac{3}{8\pi}} M_{pl} \Gamma_m  \sqrt{\Phi_m} a^{3/2},
 \eeq 
 where the last approximation is valid for $a\gg 1$.
This can be integrated to give
 \beq
 \Phi_R(a) = \Phi_{m,i} + \frac{2}{5}  \sqrt{\frac{3}{8\pi}}  M_{pl} \Gamma_m \sqrt{\Phi_{m,i}}  \lp a^{\frac 52} -1 \rp.
 \label{eq:Ra}
 \eeq
The use of approximate expression in \eqref{eq:dPhiRda} induces a negligibly small error, since in our case $M_{pl}\Gamma_m \sqrt{\Phi_m}\ll 1$. From \eqref{eq:Ra} we see that 
 for  $a\sim {\mathcal O}(a_{\rm eq}= 1)$ the $\Phi_R$ is constant, $\Phi_R(a)\simeq \Phi_{m,i}$, as expected for adiabatic expansion. Only once the second term in \eqref{eq:Ra} is comparable in size to $\Phi_{m,i}$
 does the evolution enter a non-adiabatic phase. We equate the two terms to define the transition temperature, $T_{\rm NA}$. For $a_{\rm NA}\gg a_{\rm eq}=1$, this is given by 
\beq
T_{\rm NA} = T_{\rm eq} \lp \frac{g_*(T_{\rm eq})}{g_*(T_{\rm NA})}\rp^{\frac 13} \lp\frac{2}{5}  \sqrt{\frac{3}{8\pi}} \frac{M_{pl} \Gamma_m}{\sqrt{\Phi_{m,i}}} \rp^{\frac25} 
\approx 0.59 \lp \frac{g_*^{\frac1{4}}(T_{\rm eq})}{g_*(T_{\rm NA})} \rp^{\frac13}  \lp M_{pl} \Gamma_m \Phi_{m,i}^{\frac 18}\rp^{\frac25},
\label{eq:TNA}
\eeq
where we used that up to this point the evolution is adiabatic, and therefore $g_*(T_{\rm eq}) T_{\rm eq}^3 a_{\rm eq}^3 =g_*(T_{\rm NA})T_{\rm NA}^3 a_{\rm NA}^3$. 
 During most of the NA evolution, the temperature evolves as 
\beq
\Phi_R \sim g_* T^4 a^4  \sim a^{5/2} \; \Rightarrow\; T\propto g_*^{-\frac14} a^{-\frac38},
\label{eq:Ta38}
\eeq 
as can be seen from \eqref{eq:Ra}. This $T\propto a^{-\frac38}$ scaling is utilized in derivations throughout this work. In particular, the cooling of the SM plasma with growing $a$ is slower than for the thermally decoupled HS plasma, for which the temperature is $\propto a^{-1}$. 

Once enough of the moduli decay the Hubble expansion takes over when $\Gamma_m \sim H$, and adiabatic expansion resumes.  This transition defines the reheat temperature of the universe, for which we use the standard definition \cite{Kofman:1997yn}, 
\beq
T_{\rm RH} = \lp \frac{90}{8\pi^3 g_*(T_{\rm RH})}\rp^{\frac14} \sqrt{\Gamma_m M_{pl}} \approx 0.78  g_*^{-\frac14}(T_{\rm RH}) \sqrt{\Gamma_m M_{pl}} .
\label{eq:TRH}
\eeq 
The three temperatures, $T_{\rm eq}$, $T_{\rm NA}$, and $T_{\rm RH}$, divide the early universe into the four epochs, see also Ref.~\cite{Co:2015pka} and Fig.~\ref{fig:schematicplot}. The three temperature scales can be easily related to one another
\beq
T_{\rm eq} = 6.83 \frac{g_*(T_{\rm NA})^{5/3}}{g_*(T_{\rm eq})^{2/3}g_*(T_{\rm RH})} \frac{T_{\rm NA}^5}{T_{\rm RH}^4}.
\eeq
 
The Hubble parameter during the four epochs can be approximately expressed as, 
 \beq
H(T) \approx \frac{1}{M_{\rm{pl}}} \sqrt{\frac{4\pi^3}{45}} \left\{
\begin{matrix*}[l]
 \sqrt{g_*(T)} T^2 &  \text{ ERD}, 
 \\
    \sqrt{\frac{g_*(T_{\rm NA}) g_*(T)}{g_*(T_{\rm RH})}}    \frac{T_{\rm NA}^{5/2}}{T_{\rm RH}^2}   T^{3/2}
  &  \text{ MD$_{\rm A}$}, \\
 \frac{g_*(T)}{ \sqrt{g_*(T_{\rm RH})}} \frac{T^4}{T_{\rm RH}^2} & \text{ MD$_{\rm NA}$}, \\
  \sqrt{g_*(T)} T^2 &  \text{ RD}, \\
\end{matrix*}
\right. 
\label{eq:Hubble}
\eeq
which correctly captures the temperature scalings, but neglects $\order{1}$ factors and $g_*$ ratios that can enter at interfaces between the epochs.

 \subsection{A diluted hot relic}
\label{sec:HSdil}

We turn next to our case of a plasma that is composed of both SM and HS particles.
Throughout this work we will use symbols with (without) a tilde to denote quantities in the hidden (SM) sector. 
The early cosmology of the model contains three separate energy densities: of the SM sector, $\rho$, of the hidden sector, $\tilde \rho$, and of the moduli, $\rho_{\rm mod}$.  In general, the temperature of the SM plasma, $T$, will differ from the hidden sector plasma temperature, $\tilde T$. The total radiation density is thus 
\beq
\rho_R=\rho+\tilde \rho\propto g_* T^4+\tilde g_*\tilde T^4, 
\eeq
where $g_* (\tilde g_*)$ is the number of effective relativistic d.o.f.~in the SM (HS) plasma defined with respect to that sector's temperature.  

For the dilution mechanism to work there are two important ingredients: (1) the moduli decays almost exclusively into the SM particles, which slows down the cooling of the SM sector relative to the HS, and (2) the HS plasma decouples from the SM at some time $t_D$ before $t_{\rm RH}$ (in Fig.~\ref{fig:schematicplot}, for example,  we are assuming $t_{D}$ earlier than $t_{\rm NA}$).  Beyond these two essential ingredients there are many moving parts for any particular particle physics model that realizes the dilution mechanism. To shorten the discussion, we make a few further simplifying assumptions, which, except where noted, are in place to streamline the calculations, but introduce only mild qualitative changes.

First of all, we impose the more stringent requirement that the moduli decays exclusively into SM particles, so that the hidden sector is not heated at all from the moduli decay.  
If this is not the case, there is maximum possible dilution imposed by BR(mod$\to$HS)/BR(mod$\to$SM).
We also assume that the HS remains relativistic throughout the dilution period, so that the co-moving energy densities in the SM and HS plasmas are given by $\Phi_{R,{\rm SM}}\equiv \rho a^4$ and $\Phi_{R,{\rm hs}}\equiv \tilde \rho a^4$, respectively.    In particular, we assume that $T_{\rm RH}>m_\chi$, which is the parameter range we will use in the phenomenological analysis.  Furthermore, we assume that throughout the dilution period $\rho \gg \tilde \rho$, so that to a very good approximation the Hubble rate is given by the expressions using only the SM plasma, as in Section  \ref{sec:MDU}. We also assume that the energy transfer between the SM and the HS is rapid at early times, and then abruptly shuts off below the decoupling temperature, $T_D$. The Boltzmann equations are then given by,
\beqa
\dot\Phi_m &=& - \Gamma_{m} \Phi_m\,, \label{eq:sepdiffeqbegin}\\
\dot \Phi_{R,{\rm SM}} &=& a \Gamma_{m} \Phi_m  \frac{g_{*}}{g_{*} +\tilde g_{*}\Theta\lp T - T_{D}\rp}\,,\\
\dot \Phi_{R,{\rm hs}} &=& a \Gamma_{m} \Phi_m \frac{\tilde g_{*}}{g_{*} +\tilde g_{*}}\Theta\lp T - T_{D} \rp.
\label{eq:sepdiffeqend}
\eeqa
At temperatures lower than $T_{D}$ there is no longer an efficient energy transfer between the SM and HS, $\Gamma_{{\rm hs} \leftrightarrow {\rm sm}}<H$. The instantaneous approximation for this transition is encoded by the step function, $\Theta(T-T_D)$.
The use of the step function to impose an instantaneous transition is a reasonable approximation for the models with heavy mediators. Below the mass of the mediator the evolution of the energy transfer collision term $C_E$ scales with a high power of the temperature, e.g.,~for a heavy vector mediator $C_E \sim T^9$.
We use the differential equations \eqref{eq:sepdiffeqbegin}-\eqref{eq:sepdiffeqend} along with the equation for the Hubble rate, Eq.~\eqref{eq:Hub}, to determine the cosmological history for the three sectors, $\Phi_m$, $\Phi_{R,{\rm SM}}$, $\Phi_{R,{\rm hs}}$.

For temperatures below $T_D$ the injection of entropy from the moduli decays contributes only to the SM energy density. 
We can thus define a dilution factor, 
\beq
 \label{eq:Ddef}
D\left(t\right) \equiv  \frac{s(t)}{\tilde{s}(t)} = \frac{g_{*S}(T)}{\tilde g_{*S}(\tilde T)} \lp\frac{T}{\tilde T}\rp^3,
\eeq
where $T$ and $\tilde T$ are the SM and hidden sector temperatures defined at a common late time scale factor, $s,\tilde s$ are the entropy densities of the two sectors, and $g_{*S}(T), \tilde g_{*S}(\tilde T)$ count the effective number of massless d.o.f.~in the expressions for the entropy density (in general these could differ from $g_*(T), \tilde g_*(\tilde T)$ if the SM and/or the HS are composed of more than two plasmas with differing temperatures, e.g., in the SM this happens once the neutrinos decouple from the photon plasma).  The dilution factor $D(t)$ tracks the relative entropy change between the SM and HS plasmas.  We assume that at late times entropy is conserved in the HS, so that $\tilde g_{*S}(\tilde T)\tilde T^3 a^3 = $ const. Finally, to simplify the expressions we assume that at $t>t_{\rm RH}$, the hidden sector contains only the DM particles.

As we are most interested in the total dilution, we define
\beq
D \equiv D(t_{0}), 
\eeq
where $t_0$ is the present time. Note that for $t\gg t_{\rm RH}$ the dilution factor $D(t)$ is typically constant, and equal to $D$.
Requiring that DM 
produces the correct relic abundance results in a proportionality relation between the DM mass and the required dilution factor, 
 \beq
 \label{eq:mvsD}
 m_{\chi} = \Omega_{\chi}\frac{\rho_{c,0}}{s_0} \frac{s(t_{\rm RH})}{n_{\chi}(t_{\rm RH})} = \Omega_{\chi} \frac{\rho_{c,0}}{s_0} \frac{2 \pi^4}{45 \zeta_3} \eta D = (1.5 \mbox{ eV}) \times \eta D,
 \eeq
 where $\Omega_{\chi} \simeq 0.258$ is the DM energy density fraction, $\rho_{c,0}\simeq 3.8\times 10^{-11}$ eV$^4$ the critical density, and $s_0\simeq 2.3\times 10^{-11}$ eV$^3$ the entropy density of the universe today, while the numerical factor $\eta = g_{\chi}/\tilde g_{*S}(t_{0}) =  7/6\,(1)$ for fermions (bosons) with $g_{\chi}$ is the DM number of degrees of freedom. In the first equality, we have used the fact that after $t_{\rm RH}$ the co-moving number density of DM particles is conserved, so that $n_\chi \propto 1/a^3$. In the second equality we used the definition of the dilution factor in Eq.~\eqref{eq:Ddef} to trade $s$ for $D \tilde s$, as well as the relation $\tilde s/n_\chi=\eta (2\pi^4)/(45 \zeta_3)$, valid if DM is the only d.o.f.~left in the HS after $t_{\rm RH}$ (the results are straightforward to adjust if this is not the case).
 
 Eq.~\eqref{eq:mvsD} offers an intuitive understanding of the dilution factor $D$.  The relic abundance is proportional to $\Omega_\chi \propto m_\chi/D$. That is, for DM that decouples from the SM when the DM is still relativistic, where the co-moving number density is constant, the relic abundance is bigger the greater the DM mass. The relic abundance gets diluted by the relative amount of entropy in the SM relative to the HS, i.e., by the dilution factor $D$. For adiabatic evolution the dilution factor is given simply by the ratio of effective relativistic d.o.f., $D=g_{*S}(T_D)/\tilde g_{*S}(T_D)$. This can be quite large if there are only a few relativistic d.o.f.~in the HS, since the SM contains many relativistic d.o.f.. For instance, for HS composed just of DM, the dilution factor even in the case of adiabatic evolution can be $D\sim {\mathcal O}(10-100)$. For the non-adiabatic evolution induced by the moduli decay the dilution factor can be significantly bigger, which is one of the primary points of this paper. The rest of this section is devoted to understanding the possible sizes of the dilution factor in the presence of moduli decays.
 
 As the first step, we derive the expression for the dilution factor that follows from Boltzmann equations (\ref{eq:sepdiffeqbegin})-(\ref{eq:sepdiffeqend}). Initially, we work within the approximation that there are no collisions between the SM and the HS particles. 
 During the NA period, the SM temperature evolves as $T\propto g_*^{-\frac 14} a^{-\frac 38}$, Eq.~\eqref{eq:Ta38}, while the decoupled HS evolves adiabatically, $\tilde{T}\propto \tilde{g}_{*S}^{-\frac 13} a^{-1}$. Since the HS cools more quickly this induces a dilution factor,  
 \beq
 \label{eq:D:long}
 D =  \left\{
 \begin{matrix*}[l]
  \frac{g_{*S}(T_{\rm RH})}{\tilde g_{*S}(T_{D})} \frac{g_{*S}(T_D)}{g_{*S}(T_{\rm NA})} \lp\frac{g_{*}(T_{\rm NA})}{g_{*}(T_{\rm RH})}\rp^2 \lp\frac{T_{\rm NA}}{T_{\rm RH}}\rp^5 & :~~T_D > T_{\rm NA}, \\
 \frac{g_{*S}(T_{\rm RH})}{\tilde g_{*S}(T_D)} \lp\frac{g_{*}(T_D)}{g_{*}(T_{\rm RH})}\rp^2\lp\frac{T_D}{T_{\rm RH}}\rp^5 & :~~T_D < T_{\rm NA}.
 \end{matrix*}
 \right.
 \eeq 
 Note that $D$ depends implicitly on $\Phi_m$ and $\Gamma_m$ through the value of $T_{\rm NA}$, cf. Eq.~\eqref{eq:TNA}.
 
 If $g_{*}= g_{*S}$ and $\tilde g_{*}= \tilde g_{*S}$, which is the case if both the SM and the HS sector are controlled by a single temperature each,  $T$ and $\tilde T$, then Eq.~\eqref{eq:D:long} can be shortened to,
\beq
 D =  \left\{
 \begin{matrix*}[l]
\label{eq:dil}
 \frac{g_{*}(T_D)}{\tilde g_{*}(T_D)} \frac{g_{*}(T_{\rm NA})}{g_{*}(T_{\rm RH})} \lp\frac{T_{\rm NA}}{T_{\rm RH}}\rp^5 & :~~T_D > T_{\rm NA}, \\
 \frac{g_{*}(T_D)}{\tilde g_{*}(T_D)} \frac{g_{*}(T_D)}{g_{*}(T_{\rm RH})} \lp\frac{T_D}{T_{\rm RH}}\rp^5 & :~~T_D < T_{\rm NA}.
 \end{matrix*}
 \right.
 \eeq
 The assumption $g_{*S}= g_{*}$ is true for the SM above 1 MeV, and thus true above the values of $T_{\rm RH}$ considered in this work.
For the rest of this work, we will therefore assume $g_{*S}= g_{*}$ and $\tilde g_{*}= \tilde g_{*S}$ for simplicity.
 
 \subsection{Maximum dilution due to leak-in}
\label{sec:HSdil}

The derivations in the previous section assumed that the HS can get arbitrarily cold relative to the SM plasma once the two sectors have decoupled. This is not entirely true because the residual coupling between the SM and the HS can lead to the heating of the cold HS from the much hotter SM sector via leak-in \cite{Evans:2019vxr}. This results in a \emph{lower bound} on how cold the HS can get with respect to the SM. For a given decoupling temperature $T_{D}$, there is an upper bound on the size of the entropy dilution $D$, and, from Eq.~\eqref{eq:mvsD}, an upper bound on the DM mass that is consistent with the dilution mechanism. This upper bound appears because an attempt to increase $\Phi_m$ increases the Hubble expansion rate relative to $T$. In order to maintain a fixed decoupling temperature the interaction strength between the SM and the HS (in the  case  of vector portal, the product $\alpha_D \epsilon^2$) must therefore be increased.  However, the increased coupling between the SM and the HS sectors also causes more energy to be injected into the HS at late times, which decreases the dilution.

 To derive the upper bound on $D$, consider the evolution of the HS energy density
\begin{align}
\frac{d\tilde \rho}{dt} = -4H\tilde\rho \,+ C_E(T ,\tilde T)\,.
\end{align}
The first term on the right hand side tracks the red-shifting of energy density in radiation. The second term is the energy collision term transferring energy between the two sectors.  In the $\tilde T\ll T$ approximation, generically valid when examining whether the dilution has been saturated, the energy collision term depends only on the SM temperature,
\beq
\label{eq:C_E}
C_E(T,\tilde T) \approx C_E(T) = c_E  T^{5+n}/M^n\,.
\eeq
The scaling power $n$ and the prefactor $c_E/M^n$, with $M$ a dimension-full mass parameter and $c_E$ a dimensionless factor, depend on the particular model in question. For instance,  a light axion-like particle (ALP), heavy vector, and heavy ALP have $n=2,4,6$ and $M^n=f_a^2, m_V^4, f_a^2 m_a^4$, respectively, see Table \ref{tab:scale}. 

For the HS to redshift as decoupled radiation, and therefore for the dilution to be effective, one needs $4H\tilde\rho \geq C_E (T)$. Using $\tilde \rho = ({\pi^{2}}/{30})\tilde g_{*}\tilde T^{4}$ this translates to,
\begin{align}
\label{eq:maxratio}
\frac{T^{4}}{\tilde T^{4}} \leq \frac{2\pi^{2}}{15} \frac{M^n}{c_E}\frac{\tilde g_{*} H(T) }{ T^{n+1}}\,.
\end{align}
In order to find the maximum entropy dilution we need to understand how the right hand side of \eqref{eq:maxratio} behaves as a function of $T$ after decoupling.  At decoupling we have
\beq
 4\tilde\rho(T_D)H(T_D) \approx C_E(T_{D})  \Rightarrow \frac{2\pi^{2}}{15}\frac{M^n}{c_E}\frac{\tilde g_{*}(T_D) H(T_D) }{ T_D^{n+1}}\approx 1,
 \label{eq:decouplingcond}
\eeq
 which relates the collision term parameters to the decoupling temperature. 
 
If $T_D<  T_{\rm NA}$ there is less entropy dilution than  when $T_{D} \geq T_{\rm NA}$. 
For the remainder of the derivation we therefore assume that $T_{D} \geq T_{\rm NA}$, while the analogous derivation for $T_D<  T_{\rm NA}$ is straightforward.  For $T\geq T_{\rm NA}$ we have $H \propto  g_{*}^{\frac12}(T) T^{3/2}$, cf. Eq.~\eqref{eq:Hubble}. This gives that at $T_{\rm NA}$
\beq
\left.\frac{2\pi^{2}}{15} \frac{M^n}{c_E}\frac{\tilde g_{*} H(T) }{ T^{n+1}}\right|_{T_{\rm NA}}=\frac{\tilde g_{*}(\tilde T_{\rm NA})}{\tilde g_{*}(\tilde T_D)} \left(\frac{g_{*}(T_{\rm NA})}{g_{*}(T_{D})}\right)^{1/2}\left(\frac{T_{\rm NA}}{T_{D}}\right)^{3/2} \left(\frac{T_{D}}{T_{\rm NA}}\right)^{n+1}\,.
\eeq
In the non-adiabatic regime, $T_{\rm NA}> T >T_{\rm RH} $, the SM  energy density $\rho$  heats up due to the entropy injection from the decaying moduli, giving $H \propto g_{*}(T) T^{4}$, cf.~Eq.~\eqref{eq:Hubble}. Thus, finally
\beq
\label{eq:HNSV}
\left. \frac{2\pi^{2}}{15} \frac{M^n}{c_E}\frac{\tilde g_{*} H(T) }{ T^{n+1}}\right|_{T_{\rm RH}}=\frac{\tilde g_{*}(\tilde T_{\rm RH})}{\tilde g_{*}(\tilde T_D)} \frac{g_{*}(T_{\rm RH})}{g^{1/2}_{*}(T_{D})g^{1/2}_{*}(T_{\rm NA})}\frac{T_{D}^{n-\frac{1}{2}} T_{\rm RH}^{3-n}}{T_{\rm NA}^{5/2}}\,.
\eeq

Using \eqref{eq:HNSV} in \eqref{eq:maxratio}, setting $T = T_{\rm RH}$, and utilizing $D =g_{*}(T_{\rm RH}){T^{3}_{\rm RH}}/{\tilde g_{*}(\tilde T_{RH})}{\tilde T^{3}_{\rm RH}}$,  give the upper bound
\beq
\label{eq:maxdil}
D \leq  \frac{g^{7/4}_{*}(T_{\rm RH})}{\tilde g^{1/4}_{*}(\tilde T_{\rm RH})\tilde g^{3/4}_{*}(\tilde T_{D})g^{3/8}_{*}(T_{D})g^{3/8}_{*}(T_{\rm NA})} \frac{T_{\rm RH}^{\frac{9-3n}{4}} T_{D}^{\frac{3}{4}(n-\frac{1}{2})}}{T_{\rm NA}^{15/8}}\,.
\eeq
Note that if $T_{\rm NA}$ increases, the maximal achievable $D$ decreases for a given $T_{D}$, i.e., for a given collision strength $c_E$ in \eqref{eq:C_E}. The scaling with $T_{\rm NA}$ in Eq.~\eqref{eq:maxdil} is qualitatively different than in the case of negligible SM--HS collisions, Eq.~\eqref{eq:dil}. If the SM--HS collisions can be neglected the dilution grows with $T_{\rm NA}$,  $D\propto T_{\rm NA}^5$. In contrast, when the collisions are important $D$ decreases with $T_{\rm NA}$, $D\propto T_{\rm NA}^{-15/8}$. 
This implies that increasing $T_{\rm NA}$, while keeping $T_{D}$ and $T_{\rm RH}$ fixed, will cause the entropy dilution factor $D$ to grow, as in Eq.~\eqref{eq:dil}, up to its maximal value, $\bar D_{\rm max}$,  
after which it starts to decrease. 
By equating Eqs.~\eqref{eq:dil} and \eqref{eq:maxdil}, we can determine the optimal value of $T_{\rm NA}$ that corresponds to the maximal entropy dilution,
\begin{align}
\label{eq:TNAmax}
\left. T_{\rm NA}\right|_{\bar D_{\rm max}} = \left(\frac{\tilde g_{*}(T_D)}{ \tilde g_{*}(\tilde T_{\rm RH})}\right)^{\frac{2}{55}} \frac{g^{2/5}_{*}(T_{\rm RH})}{g^{1/5}_{*}(T_{D})g^{1/5}_{*}(T_{\rm NA})}  \lp \frac{T_D}{T_{\rm RH}}\rp^{\frac{6n-3}{55}} T_{\rm RH}\,.
\end{align}
Using this in Eq.~\eqref{eq:dil} gives one of the main results of this paper, the maximum dilution assuming internal thermalization in the HS sector, 
\begin{align}
\label{eq:Dmax}
\bar D_{\rm max} = \frac{g_{*}(T_{\rm RH})}{\tilde g^{9/11}_{*}(\tilde T_D)\tilde g^{2/11}_{*}(\tilde T_{\rm RH})}\left( \frac{T_D}{T_{\rm RH}} \right)^{\frac{3 (2n -1)}{11}} \equiv \bar \lambda_D \left( \frac{T_D}{T_{\rm RH}} \right)^{\bar \gamma_n},
\end{align}
where we have defined the $g_*$ ratios to be $\bar \lambda_D$.  The scaling parameter $\bar \gamma_n \equiv 3 (2n -1)/11$ depends on the mediator model, and is $\bar \gamma_n=9/11, 21/11, 3,$ for light ALP, heavy vector and heavy ALP, respectively, see Table \ref{tab:scale}.

Using the maximal dilution factor $\bar D_{\rm max}$ in Eq.~\eqref{eq:mvsD} gives
\beq
\left. T_D\right|_{\bar D_{\rm max}} \simeq \frac{1}{\bar \lambda_D^{1/{\bar \gamma_n}}} \lp \frac{m_{\chi}}{1~{\rm eV}} \rp^{\frac{1}{\bar \gamma_n}} T_{\rm RH}\,.
\eeq
This is the smallest $T_D$ for which there exists a dilution model consistent with DM mass $m_\chi$. 
We typically expect $\bar \lambda_D\sim {\mathcal O}(10)$, while $\bar \gamma_n$ typically spans values ${\mathcal O}(1-\text{few})$. For DM masses well above an eV to be viable therefore requires a large hierarchy between $T_D$ and $T_{\rm RH}$. For instance, for a heavy vector mediator $\bar \gamma_n=21/11\sim 2$. For DM mass $m_\chi \sim {\cal O}(1)$ MeV to be viable we need in this case $T_D/T_{\rm RH}\sim10^3$. Thus, if we allow the reheating temperature to be close to the Big Bang Nucleosynthesis (BBN) limit of $T_{\rm RH}\sim 2$~MeV (discussed in detail in Section \ref{sec:BBN}), the decoupling temperature needs to be at least $T_D \sim 1$~GeV.

\begin{table}
\center{\begin{tabular}{ccccc}
\hline\hline
 \mbox{ } Example Model\mbox{ }  & \mbox{~~}$n$\mbox{~~} & \mbox{~~}$\bar \gamma_n$\mbox{~~} & \mbox{~~}$\gamma_n$\mbox{~~} & $M^n$\\
\hline
 Light ALP &2 & $\frac{9}{11}$  & $1$  & $f_a^{2}$ \\
 Heavy Vector&4 & $\frac{21}{11}$&  $\frac{7}{3}$  & $m_V^{4}$\\
Heavy ALP & 6 & 3  &  $\frac{11}{3}$  & \mbox{ } $f_a^{2}m_a^{4}$\mbox{ } \\
\hline\hline
\end{tabular}}
\begin{caption}{
The scaling powers $\bar  \gamma_n$ ($\gamma_n$) for the maximal dilution factor with (without) internal thermalization of the dark sector  for three different mediator models, see also Eqs.~\eqref{eq:C_E}, \eqref{eq:Dmax}, \eqref{eq:Dmax:IT}.
\label{tab:scale}
}
\end{caption}
\end{table}

So far we assumed that the HS maintains internal thermalization throughout the cosmological evolution. Any energy injected into the HS then heats the HS and results in a thermal distribution with a new, higher temperature $\tilde T$. Typically, the assumption of complete internal thermalization requires sizable coupling of the mediator to the HS particles (in the case of dark photon the value of $\alpha_{D}$), often in conflict with experimental searches.  
Alternatively, the interactions among the HS and the interactions between the SM and the HS can due to different mediators. In this way 
internal thermalization in the HS is possible without too much impact on the HS--SM phenomenology. 

Even if internal thermalization is not maintained, the injection of energy into the SM plasma and the subsequent energy transfer to the HS, either through pair creation of the HS particles or via collisional energy transfer to the HS particles, still places an upper bound on the allowable dilution.  In this case, the injected DM particles simply redshift the excess energy away rather than being converted into multiple DM particles. A detailed discussion is given in Appendix \ref{app:IT}. Here we only quote the result for the extreme case of no internal thermalization in the HS throughout the relevant temperature range, 
\beq
\label{eq:Dmax:IT}
D_{\rm max}= \lp\frac{45 \zeta_3 }{2\pi^4 \kappa\eta}\rp^{\frac{2}{3}} \frac{g_*(T_{\rm RH})}{\tilde g_*(T_{D})} \lp\frac{T_D}{T_{\rm RH}} \rp^{\frac{2n-1}{3}}\equiv\lambda_D \lp\frac{T_D}{T_{\rm RH}} \rp^{\gamma_n}.
\eeq  
Here $\kappa \equiv T C(T, \tilde T)/C_{E}(T, \tilde T) \sim \mathcal{O}(1)$, while $\gamma_n$ is the modified scaling power, see Table \ref{tab:scale}. 
For our benchmark model presented in Section \ref{sec:vector}, the sector is not internally thermalized, so that \eqref{eq:Dmax:IT} gives the maximum achievable dilution. 
Note that DM masses well above eV require large $T_D/T_{\rm RH}$ ratios also in the case when the HS does not thermalize. 

\section{Model Independent Constraints on Diluted Hot Relics}
\label{sec:HS:constr}

Several important and nearly model independent constraints bound the diluted hot relic parameter space. In this section, we describe in detail the constraints  that arise from the free-streaming of DM,  the Tremaine-Gunn bounds on dwarf galaxies, and constraints from BBN. The constraints are summarized in Fig.~\ref{fig:MassBounds}, with the most stringent constraints  bounding $m_\chi >4.4$ keV and $T_{\rm RH}>2$ MeV.

\begin{figure}[t]
\center{\includegraphics[width=0.6\textwidth]{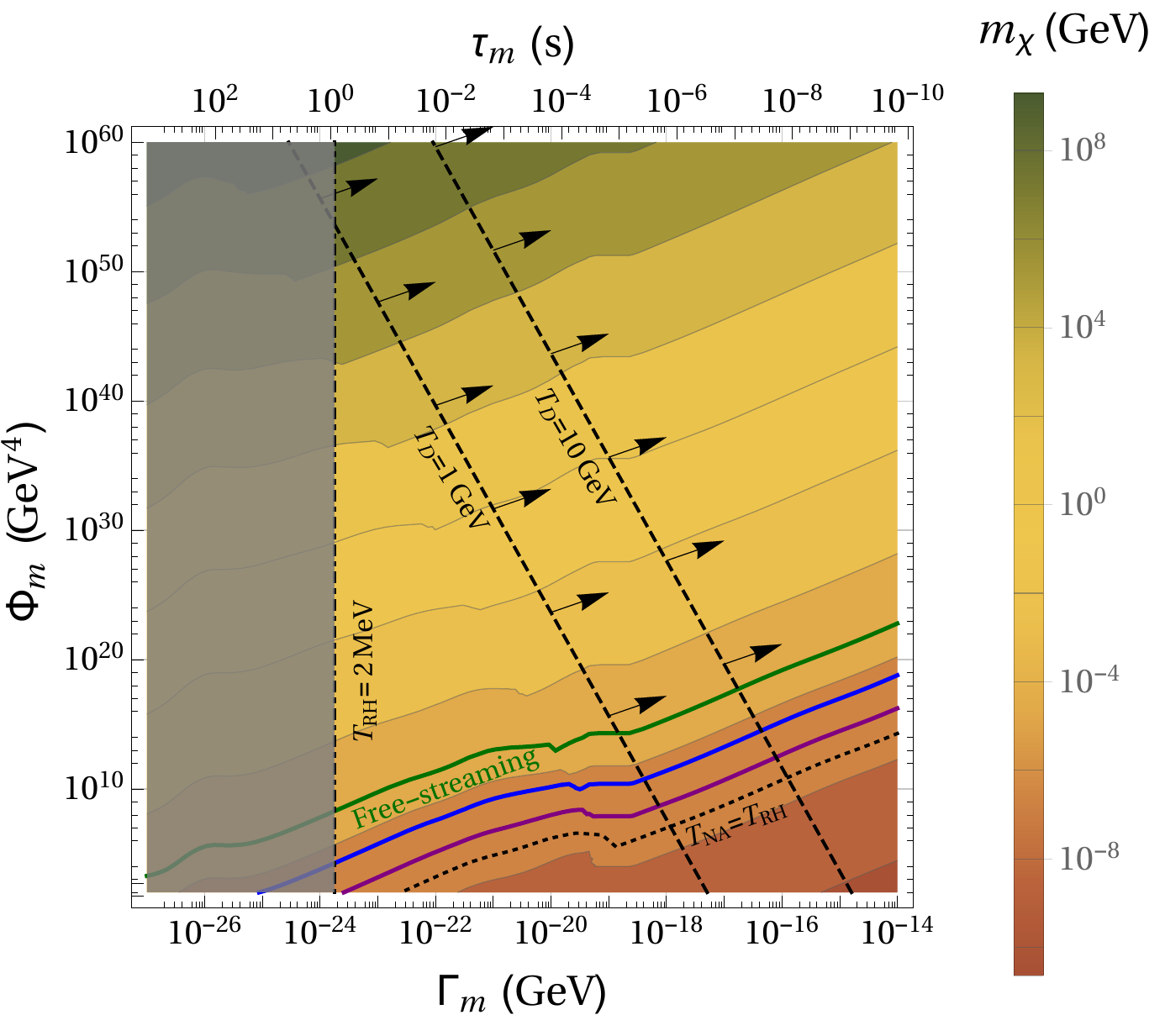}
\caption{Isocontours of DM mass, $m_{\chi}$,  in the $ \Gamma_m-\Phi_m$ plane (i.e., the decay rate vs. energy stored in the moduli), assuming $T_D > T_{\rm NA}$, and not imposing an upper bound on the dilution, so that $D$ is given by Eq.~\eqref{eq:dil}.  
The two black dashed lines are the $T_D = T_{\rm NA}$ limiting curves  for $T_D = 1 (10)$ GeV, indicating in each case that the $m_{\chi}$ contours to the right would change since the $T_D < T_{\rm NA}$ expression for $D$ needs to be used instead. 
The solid green, blue and purple lines denote the lower limits on $m_{\chi}$ from free-streaming, Tremaine-Gunn and BBN, respectively. On the dotted black line $T_{\rm RH} = T_{\rm NA}$, so that near and below this line there was no MD epoch at all. The black dot-dashed line corresponds to $T_{\rm RH} = 2$ MeV, with the grayed out region to the left excluded by BBN, see text for details. 
While this work focuses on light DM, we show the effect of dilution up to very heavy masses.
}
\label{fig:MassBounds}}
\end{figure}

\subsection{Collisionless damping (free-streaming)}
\label{sec:fs}

Free-streaming DM erases features in gravitational potentials at small scales, 
causing a suppression of the matter power spectrum on the DM free-streaming scales. 
Lyman$-\alpha$ forests \cite{Irsic:2017ixq,Viel:2004bf} trace the cosmological perturbations by looking at the absorption spectra of background quasars at redshifts $z \sim (2-4)$. Using Lyman$-\alpha$ forests, it is possible to probe cosmological perturbations at scales smaller than about $\sim 100$ kpc. 

The warm dark matter constraints derived in Ref.~\cite{Irsic:2017ixq} assume a thermal number abundance at a colder temperatures, which allows for robust bounds to be placed on a variety of models, see e.g.~\cite{Bae:2017dpt,Kamada:2019kpe}.  These bounds are typically derived for sterile neutrinos, which assumes two d.o.f.\ for the dark matter.  In our case there are four d.o.f., which relaxes the bound on the dark matter mass by $2^{1/4}$ \cite{Bode:2000gq} and brings the bound of 5.3 keV~\cite{Irsic:2017ixq} down to 4.4 keV.  

\subsection{Tremaine-Gunn constraints}

Limits on collisionless DM due to the conservation of phase-space density (PSD) from Liouville's theorem are generally referred to eponymously as Tremaine-Gunn constraints \cite{Tremaine:1979we,Madsen:1991mz}.  The basic idea underlying these limits is that after DM becomes collisionless, the microscopic PSD is conserved, while the maximum macroscopic (course-grained) PSD for a system must decrease with time \cite{Tremaine:1979we}.  The current maximum macroscopic PSD for a DM system, e.g., a dwarf galaxy, puts a lower bound on 
 the maximum \emph{microscopic} PSD at the time of DM kinetic decoupling. 
 While the coarse-grained PSD is not precisely known, it can be estimated from the halo parameters: the velocity dispersion,  $\sigma_v$,  and the half-light radius, $r_h$, which should approximately track the DM.  This allows us to bound the diluted hot relic DM mass to be above \cite{Tremaine:1979we,Madsen:1991mz,Boyarsky:2008ju}
\beq\label{eq:TGbound}
m_{\chi} \geq  \lp  
\frac{\sqrt 3 \pi \ln 2 M_{pl}^2}{2\sqrt{2} \sigma_v r_h^2} 
\frac{1}{g_\chi \,\mbox{Max}[f(\tilde T_{\rm IT})]}\rp^{1/4} \approx (0.25-0.8)\times \ \big(g_\chi \mbox{Max}[ f(\tilde T_{\rm IT})]\big)^{-1/4} \mbox{ keV},
\eeq
where $f(T) =\lp e^{E/T} + 1 \rp^{-1}$ for fermion and $f(T) =\lp e^{E/T} - 1 \rp^{-1}$ for boson DM, and $\tilde T_{\rm IT}$ is the temperature at which the HS loses internal thermal equilibrium (see Appendix~\ref{app:IT}) . The presented range illustrates the observational uncertainties, mainly due to  smaller, ultra-faint dwarfs which have larger uncertainties in the measured halo parameters.  These dwarfs may in principle place the most stringent constraints, but with rather large systematic uncertainties. It is more robust to focus on the better understood dwarfs, such as Draco, that place weaker constraints (at the level of $0.4$ in the range above).  In our case, DM is still relativistic at the time of decoupling, giving the maximum PSD at kinetic decoupling to be
\beq
\mbox{Max}[f(\tilde T_{\rm IT})] = \left\{ 
\begin{tabular}{ccl} 
$\frac12$ & & : fermions,\\
$\tilde T_{\rm IT}/m_\chi \gg 1$ & & : bosons. 
\end{tabular}
\right.
\eeq
For $g_\chi\geq 2$ the Tremaine-Gunn constraints on the DM mass are therefore always weaker than the collisionless damping constraint, Sec.~\ref{sec:fs}, even if one uses the more uncertain ultra-faint dwarfs to place the constraints. The blue line in Fig. \ref{fig:MassBounds} shows the Tremaine-Gunn bound in the case of fermion DM, considering the central value of Eq. \eqref{eq:TGbound}, which gives $m_\chi \gtrsim 0.44$ keV.

\subsection{Big Bang nucleosynthesis}
\label{sec:BBN}

Models that produce new cosmological activity at temperatures 50 keV $\lesssim T\lesssim 1$ MeV can disrupt the delicate predictions of BBN that accurately match observations~\cite{Alpher:1948ve,Alpher:1950zz,Walker:1991ap,Pospelov:2010hj,Cyburt:2015mya}. The decaying moduli sets the reheating temperature, $T_{\rm RH}$. It is essential that the moduli does not continue to inject appreciable amounts of entropy below $T\sim 1$ MeV, so as not to disrupt the primordial abundance of $^4$He.  While the bound on the reheat temperature for a standard model profile is 4.7 MeV from Planck data \cite{deSalas:2015glj}, an atypical injection profile from the decaying massive moduli can relax this bound to as low as 2 MeV \cite{Barenboim:2017ynv}.  Since we are allowing for atypical moduli decays, we will use the conservative lower bound of $T_{\rm RH}=2$ MeV in presenting the numerical results, but stress that particular models may require $T_{\rm RH}\gtrsim 5$ MeV.  Requiring $T_{\rm RH}>2$ MeV places an upper bound on the lifetime of the moduli, $\tau_m \lesssim (1.2~\text{s})\times \sqrt{g_*(T_{\rm RH})}$, see Eq.~\eqref{eq:TRH} and gray shaded region in Fig. \ref{fig:MassBounds}.  

The agreement of the standard BBN with observations also places constraints on the number of relativistic species during the BBN epoch \cite{Cyburt:2015mya}.  After electrons leave the SM plasma, the radiation energy density of the universe can be expressed as
 \beq
 \rho_R = \frac{\pi^2}{30} \lp 2 T^4 +\frac 74 N_\nu T_\nu^4  + \tilde g_* \tilde T^4   \rp = \frac{\pi^2}{30} T^4 \lp 2 +\frac 74 N_\nu \lp\frac 4{11}\rp^{4/3} \!\!\! + \tilde g_*(\xi T) \xi^4 \rp, 
  \eeq
  with $\xi\equiv \tilde T/T$.  
The effective number of neutrino species 
in the SM is equal to $N_\nu = 3.046$, where the slight increase above $N_\nu=3$ is due to residual $e^\pm$ interactions with neutrinos after decoupling \cite{Mangano:2005cc}.  Any other contributions to $\rho_R$, such as from the HS plasma, can be reinterpreted as the additional effective number of neutrinos, $\Delta N_\nu$. Detailed measurements of nuclear relic abundances and CMB data give $N_\nu=2.88\pm 0.16$~\cite{Cyburt:2015mya}, closely matching the precise predictions from BBN.  Assuming that the SM sector is minimally influenced by the HS content and moduli decay, this imposes the constraint $\Delta N_\nu < 0.15$. The bound on the HS contribution to $\Delta N_\nu$ can be converted to a bound on the temperature in the HS,  
\beq
\Delta N_\nu = \frac 47 \lp\frac {11}4 \rp^{4/3}\tilde g_*(\xi T) \xi^4  \Rightarrow  \xi < 0.51 \, \tilde g_*(\xi T)^{-1/4} \simeq 0.376,
\eeq
where in the last equality we assumed that the HS consists of a single Dirac fermion DM.  Using Eqs.~\eqref{eq:Ddef}--\eqref{eq:mvsD}, this bound can be converted to a lower limit on the DM mass,
\beq
m_{\chi} > 0.1 \mbox{ keV}.
\eeq
This is quite less stringent than the collisionless damping constraint in Sec.~\ref{sec:fs}.

\section{A Vector Portal Model for Diluted Hot Relic Dark Matter}
\label{sec:vector}

We now apply the dilution mechanism to a specific model: the SM supplemented by a massive dark photon, $A'$, and a Dirac fermion DM, $\chi$. Many other possible DM models exist where the dilution mechanism could be effective.  
The renormalizable Higgs or neutrino portal could mediate DM interactions with the SM, while the higher-dimension axion portal is also very well-motivated.  A renormalizable vector portal for one of the anomaly-free symmetries of the SM could also be used.
As higher dimension connections between the SM and hidden sector are ultimately what allows for the dilution to be effective, a variety of other non-renormalizable operators could also be used to introduce the requisite IR decoupling.  We leave the exploration of these possibilities for future work.  

\subsection{The vector portal model}
The most general dark sector Lagrangian containing a dark $U(1)_D$ gauge boson, $\hat A'$, and Dirac fermion DM, $\chi$, is 
given by,
\beq\label{eq:KM}
\mathcal{L} \subset -\frac{1}{4} \,\hat B_{\mu\nu}\, \hat B^{\mu\nu} - \frac{1}{4} \,\hat A_{\mu\nu}'\, \hat A'{}^{\mu\nu}  + \frac{1}{2}\,\frac{\epsilon}{\cos\theta} \,\hat A'_ {\mu\nu}\,\hat B^{\mu\nu} + \frac{1}{2}\, m_{A',0}^2\, \hat A'{}^\mu \, \hat A'_{\mu}- g_D \hat A'{}^\mu(\bar \chi\gamma_\mu\chi)\, ,
\eeq
where $\theta$ is the Weinberg angle, $\epsilon$ is the kinetic
mixing parameter, $g_D$ the $U(1)_D$ gauge coupling constant, $\hat B_{\mu\nu}
=\partial_\mu \hat B_{\nu} - \partial_\nu \hat B_{\mu}$ and $\hat
A'_{\mu\nu} =\partial_\mu \hat A'_{\nu} - \partial_\nu \hat A'_{\mu}$
are the $U(1)_Y$ and
$U(1)_D$ field strengths, respectively. The hatted fields, $\hat A', \hat B$ indicate the original fields with non-canonical
kinetic terms, while we denote with $A'$ and $B$ the canonically normalized fields. The mass $m_{A',0}^2$ could originate from a St\"uckelberg \cite{Stueckelberg:1938zz,Feldman:2007wj} or a Higgs mechanism.  As long as any massive content is non-relativistic by $T_D$, and does not increase the strength of the collision term, it is irrelevant for the cosmology.\footnote{In principle, a massive Higgs state could increase the DM self-interaction and maintain internal thermalization longer, thus lowering $T_{\rm IT}$.} 
For $m_{A',0}\ll m_Z$ and $\epsilon\ll 1$ the mass of the physical dark photon and its couplings to the SM fermions and DM are approximately 
\begin{eqnarray}\label{eq:generalLightA'}
m^2_{A^\prime}&\simeq& m^2_{A',0}(1-\epsilon^2\tan^2\theta)\,,\qquad 
g_{A^\prime f\bar f}\simeq e Q \epsilon\,,\qquad
g_{A^\prime \chi\bar \chi}\simeq g_D\left(1-\tfrac{1}{2}\epsilon^2\tan^2\theta\right)\,,
\end{eqnarray}
with $Q$ the fermion electric charge in units of $e$. 

Depending on the values of $\epsilon$ and $g_D$, the dark photon predominantly decays either to DM or to the SM fermions. The corresponding partial decay widths, at the first order in $\epsilon$, are given by
\begin{eqnarray}\label{eq:partialWA'}
\Gamma(A^\prime\to\bar ff)&\simeq& \frac{\alpha_{\rm em}N_c}{3 m_{A^\prime}}\epsilon^2Q^2_f(m^2_{A^\prime}+2m_f^2)\sqrt{1-\frac{4m_f^2}{m^2_{A^\prime}}}\,,\\
\Gamma(A^\prime\to\bar \chi\chi)&\simeq&  \frac{\alpha_D}{3 m_{A^\prime}}(m^2_{A^\prime}+2m_\chi^2)\sqrt{1-\frac{4m_\chi^2}{m^2_{A^\prime}}}\,.
\end{eqnarray}
For dark photon decays into quarks, the tree-level expression \eqref{eq:partialWA'} is a good approximation only for $m_{A^\prime}$ well above the $\bar bb$ threshold. For smaller masses, threshold effects and hadronic resonances cannot be neglected. To obtain consistent predictions for the dark photon widths across the relevant parameter space, we must include experimental information. 
This is most easily achieved by constructing the ratio 
\beq
R_{A^\prime}(m_{A^\prime})\equiv \frac{\Gamma(A^\prime\to{\rm{hadrons}})}{\Gamma(A^\prime\to\mu^+\mu^-)}=\frac{\sigma(e^+e^-\to{\rm{hadrons}})}{\sigma(e^+e^-\to\mu^+\mu^-)}\,.
\eeq
This inclusive hadronic decay ratio can be extracted experimentally from $e^+e^-$ collisions \cite{Tanabashi:2018oca}.

Now we have all the ingredients to match this model to the discussion in the previous sections. For concreteness, we consider two benchmark masses, 
\beq
\label{eq:mass:benchmarks}
\textbf{~~~~~~~DM mass benchmarks:}\qquad m_{\chi} = 5 \text{~keV},\quad\text{and}\quad m_{\chi} = 100\text{~keV},
\eeq
and two benchmark dark $U(1)_D$ gauge couplings,
\beq
\label{eq:couplings:benchmarks}
\textbf{coupling benchmarks:}\qquad  \alpha_{D}  = 10^{-3},\quad\text{and}\quad\alpha_{D}  = 10^{-9}.
\eeq
In principle, we could consider heavier DM masses. However, light DM is the focus of this work for two reasons.  Firstly, there has been recent interest in methods for detecting sub-MeV DM in terrestrial experiments \cite{Battaglieri:2017aum}.  However, the canonical models of light DM are often at odds with cosmological constraints. The dilution mechanism we presented allows for light DM with sizable couplings to the SM that can fall under the purview of coming and proposed experiments.
  Secondly, for heavier DM the dilution mechanism requires larger $T_{D}$, and consequently the effective coupling between DM and the SM has to be small.  This makes the diluted hot relic DM models with heavy DM mass more difficult to detect via terrestrial experiments.   
  
To obtain accurate results in modeling the dilution as in Sec.~\ref{sec:dilution}, it is important that the two sectors decouple sufficiently below the resonance so that the collision term can be reliably treated as $C_E \sim T^9/m_{A'}^4$.  To impose this, we require that $T_D < m_{A'}/3$.

\subsection{Cosmology of the vector portal model with dilution}

The model-independent results from Section \ref{sec:dilution} are directly applicable to the vector portal model. Since we are interested in light DM, we restrict the discussion to $m_{\chi} < T_{\rm RH}$. 
For our benchmark models we choose the lowest possible value for the reheat temperature, $T_{\rm RH} = 2$ MeV, unless noted otherwise. 
This translates into the lowest possible decoupling temperature, $T_{D}$, for a given $m_{\chi}$, and thus the largest coupling between DM and the SM, maximizing the reach of the terrestrial experiments.

We define $T_D$ to be the temperature at which the Hubble expansion rate equals the DM annihilation rate into the SM particles,
\beq
H(T_D) = \Gamma_{}(T_D)=\langle \sigma_{\bar \chi \chi \to \bar f f} v\rangle n_\chi \big|_{T_D},
\eeq
which is roughly equivalent to the temperature when the collision term transferring energy between the sectors falls below the red-shifting term in the Boltzmann equation.\footnote{DM annihilation is not the only energy transfer mechanism, there is also the DM-SM scattering. The above definition of $T_D$ simplifies expressions.}
For a heavy vector with a light Dirac fermion DM, we have
\beq
\langle \sigma_{\bar \chi \chi \to \bar f f} v\rangle n_\chi \big|_{T_D}\simeq \frac{4\zeta(3)}{\pi}g_\chi \alpha_{\rm em}\epsilon^{2} \alpha_{D} \frac{T^{5}_{D}}{m_{A'}^4} \sum_{m_f<T_D} Q_f^2,
\eeq
where the sum is over the SM fermions of charge $Q_f$ that are light enough to be produced in the typical collision at temperature $T_D$.
Using the Hubble rate for expansion during the MD epoch in \eqref{eq:Hubble}, gives  
\beq
\label{eq:TD:alphaDeps}
m_{A'}^4  \propto\alpha_{\rm em}\epsilon^{2} \alpha_{D} {M_{pl} T_{\rm RH}^2 T_D}.
\eeq
That is, 
an increased decoupling temperature requires either larger $m_{A'}$ or smaller $\epsilon^2 \alpha_D$, since either of the two make the coupling of the HS to the SM weaker.  

Fig.~\ref{fig:BRApAndLifeTime} illustrates the properties of the dark photon that yield the correct value of $D$ such that the observed DM density is obtained, assuming maximal dilution. 
The mass of a diluted hot relic DM is directly proportional to the dilution factor $D$, cf.~Eq.~\eqref{eq:mvsD}. Fig.~\ref{fig:BRApAndLifeTime} left (right) gives the results for $m_\chi=1\,(100)$ keV and thus for $D\simeq 570~(57000)$.  In Fig.~\ref{fig:BRApAndLifeTime} the reheat temperature is fixed to $T_{\rm RH}=2$ MeV, which together with $D=D_{\rm max}$ from Eq.~\eqref{eq:Dmax:IT} determines $T_D$. 
This value of $T_D$ in turn determines the combination ${m_{A'}^4}/{\epsilon^{2} \alpha_{D}}$, cf. Eq.~\eqref{eq:TD:alphaDeps}. 
For a given $m_\chi$ and $T_{\rm RH}$, the dark photon coupling constant $\alpha_D$ is therefore determined everywhere in the $m_{A'}$~vs.~$\epsilon$ parameter space.  
 Note that in the parameter space we consider, $\alpha_{D}$ is not large enough for the dark sector to maintain internal thermal equilibrium until $T_{\rm RH} = 2\,\MeV$, and thus $ D_{\rm max}$ expression from \eqref{eq:Dmax:IT} applies (see also discussion in App. \ref{app:IT}).  

The properties of the dark photon vary significantly over the viable parameter space. For larger values of $\epsilon$ the dark photon decays predominantly into visible states, for smaller values of $\epsilon$ predominantly to DM pairs. 
This is illustrated in Fig.~\ref{fig:BRApAndLifeTime} where isocontours of branching ratios of dark photon to the SM and/or HS, the dark photon decay times, and the value of $\alpha_D$ that produces the correct relic abundance for $D=D_{\rm max}$ are denoted with  black dashed, green solid, and blue solid lines, respectively (red dashed lines denote equal branching ratios for decays to the SM and the HS). The correct DM relic abundance is possible both for dark photon that decays almost exclusively into visible states, as well as for predominantly invisibly decaying dark photon. 

\begin{figure}
  \centering
  \includegraphics[width=.49\linewidth]{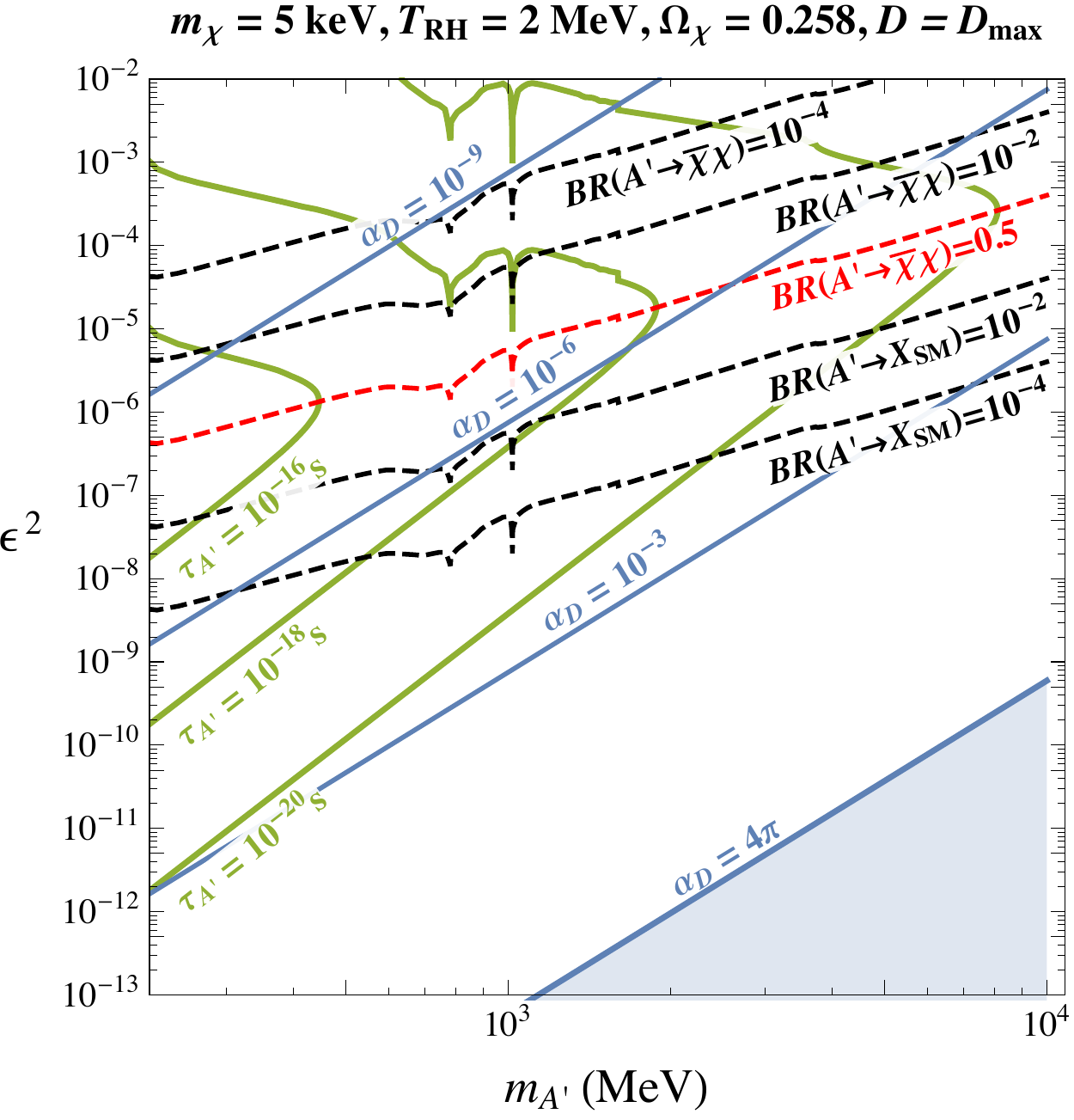}
  \includegraphics[width=.49\linewidth]{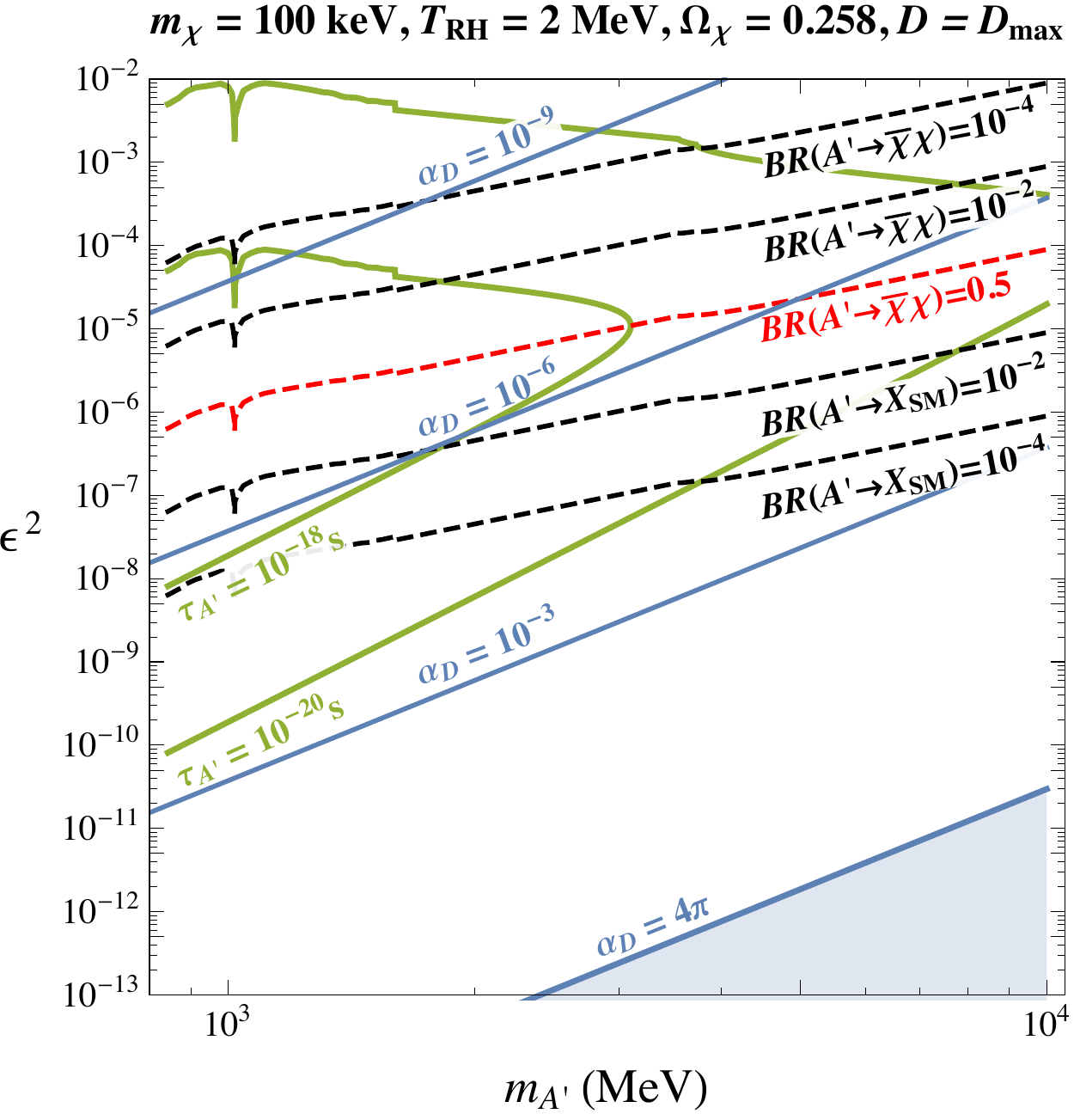}
  \caption{Isocontours of  $A'$ decay lifetimes (in green), values of $\alpha_D$ (in blue), and decay branching ratios (in black) assuming $T_{\rm RH} = 2~\MeV$ and maximum dilution, $D_{\rm max}$, for a DM of mass $m_\chi=5~\keV$ (left) and $m_\chi=100~\keV$ (right) as a function of $\epsilon$ and $m_{A^\prime}$. The red dashed line corresponds to ${\rm{BR}}(A^\prime\to\chi_{\rm{SM}})={\rm{BR}}(A^\prime\to\bar\chi\chi)$. 
  }
\label{fig:BRApAndLifeTime}
\end{figure}

 \subsection{Current and future constraints}
 There are several potential methods to probe DM.  Two of the most common, indirect detection of particles produced in DM annihilation products and direct detection of scatters off of controlled terrestrial experiments, do not apply.  Indirect detection is immensely suppressed, as the only kinematically accessible SM particles are the photon and neutrino.  However, even if the electron were accessible, the very small annihilation cross-section is still very far below sensitivity of any existing or prosed experiments.   Direct detection for very light DM masses is typically more sensitive to electron recoils when these processes exist.  However, the resulting cross-section \cite{Emken:2019tni}
 \beq
\bar \sigma_e = 16\pi \epsilon^2 \alpha_D \alpha_{\text{em}} \frac{\mu_{e\chi}^2}{m_{A'}^4} \sim \alpha_{D} \lp \frac{\epsilon}{10^{-4}}\rp^2  \lp \frac{{\text{GeV}}}{m_{A'}}\rp^4 \lp \frac{m_{\chi}}{{\text{keV}}}\rp^2 \text{ yb}
 \eeq
 is many orders of magnitude below the projected reach of even the most ambitious projects \cite{Hochberg:2019cyy, Geilhufe:2019ndy}.  Prospects for detection in nuclear recoils are even grimmer.   
  
However, there are terrestrial experiments and astrophysical observations that can place constraints on our benchmark models,  defined by Eqs.~(\ref{eq:mass:benchmarks})~and~(\ref{eq:couplings:benchmarks}). The present exclusions (shaded regions) and future constraints (dotted lines) are shown in Figs.~\ref{fig:IndepBounds}~--~\ref{fig:IndepBounds1},
as a function of the kinetic mixing parameter $\epsilon$ and the heavy mediator mass $m_{A^\prime}$. 

\begin{figure}
  \centering
  \includegraphics[width=.49\linewidth]{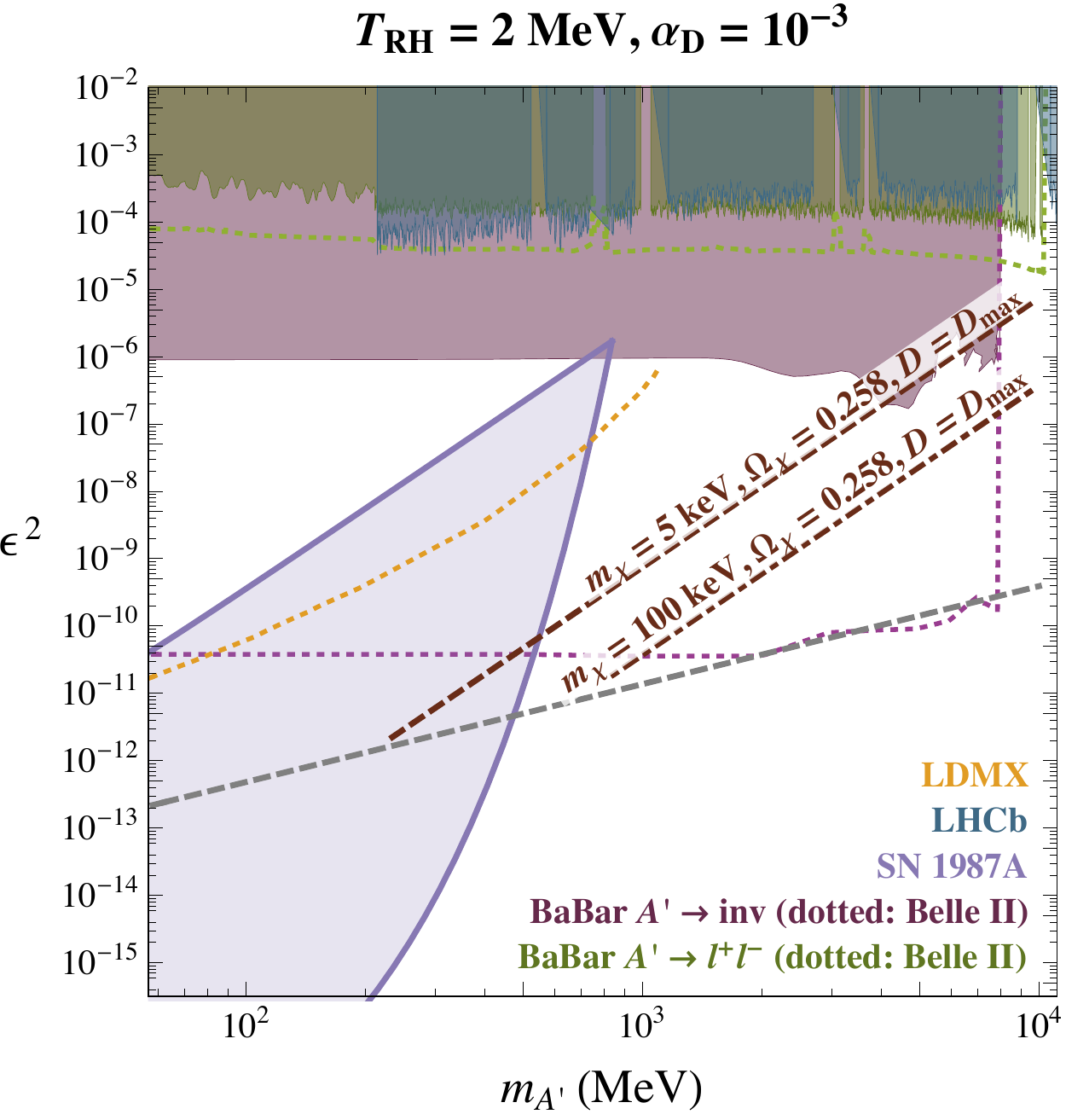}
  \includegraphics[width=.49\linewidth]{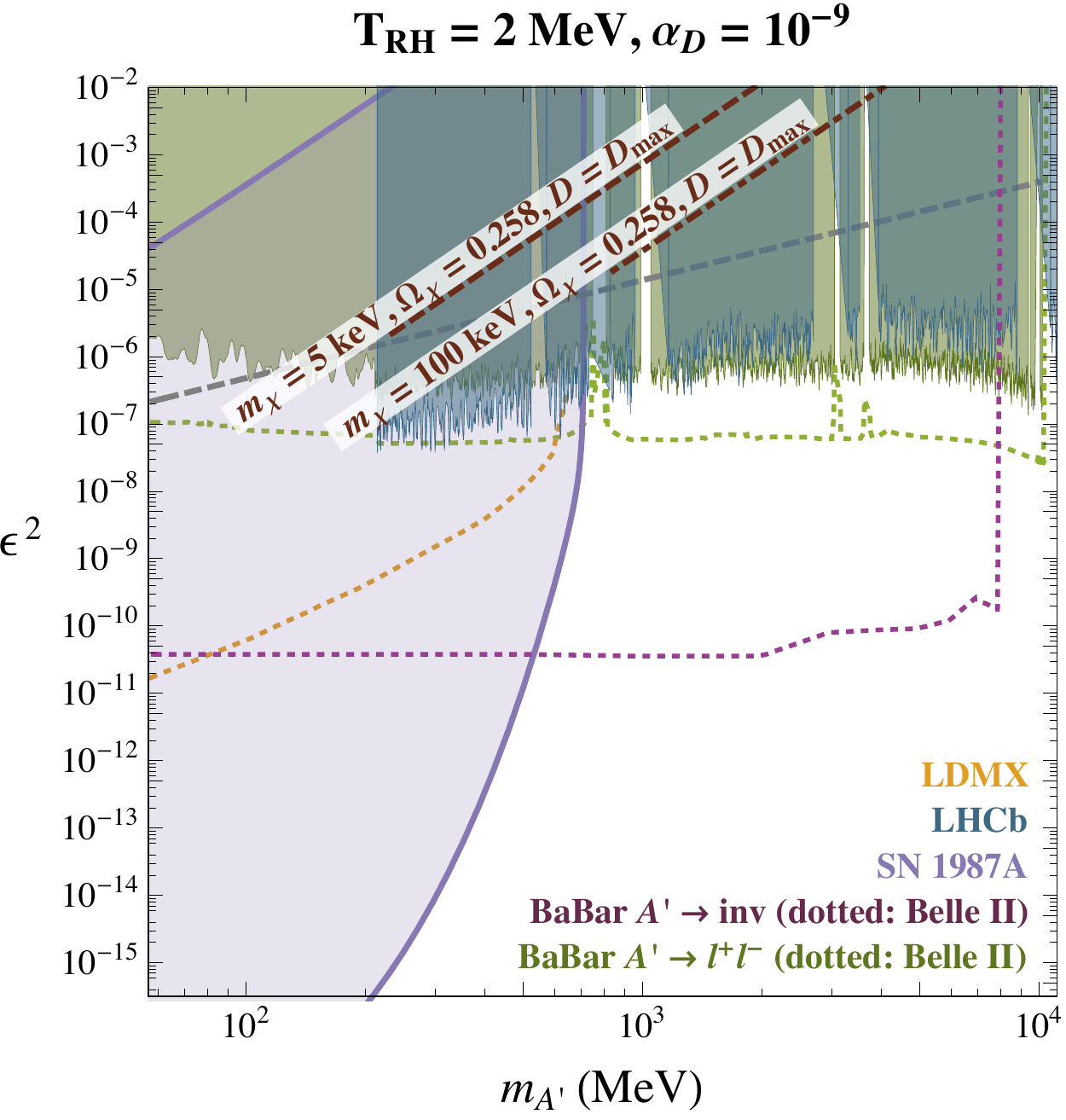}
  \caption{ 
 The present bounds on the ($m_{A'}, \epsilon^2$) vector portal parameter space for $\alpha_D=10^{-3}$ ($\alpha_D=10^{-9}$) are shown in the left (right) panel as blue, purple, maroon, and green shaded regions (from LHCb, supernovae 1987A, BaBar $A'\to$ inv, and $A'\to \ell^+\ell^-$ searches, respectively). The projected reach from Belle II is shown with maroon and green dotted lines, and from LDMX with a dotted orange line.
 The correct diluted hot relic DM abundance is obtained for $m_\chi=5$ keV ($100$ keV) on dark red dot-dashed lines, assuming maximal dilution with no internal thermalization \eqref{eq:Dmax:IT}, and reheat temperature $T_{\rm RH}=2$ MeV.  The dashed gray line below these contours denotes the limit until which our calculations of relic abundance are reliable. 
 }
\label{fig:IndepBounds}
\end{figure}

 In Fig.~\ref{fig:IndepBounds} left (right) panel the dark photon gauge coupling is fixed to $\alpha_{D}=10^{-3}~(10^{-9})$, while reheat temperature is set to $T_{\rm RH}=2$ MeV. Assuming maximal dilution, the correct diluted hot relic DM abundance is obtained  on dashed (dot-dashed) lines for $m_\chi=5$ keV ($100$ keV).  Since the internal thermalization is not maintained  by the HS during the relevant period of evolution the dilution is obtained from Eq.~\eqref{eq:Dmax:IT}.
 The gray dashed line denotes the limit until which our calculations of the hot DM relic abundance are reliable (it does not, however, imply the end of the viable parameter space leading to hot DM relic).   

In Fig.~\ref{fig:IndepBoundsRH}, the DM mass is fixed to $m_{\chi}=5$ keV, while the correct diluted hot relic DM abundance contours are calculated for sample reheat temperatures, $T_{\rm RH}=2$ and $T_{\rm RH}=10$ MeV (shown for both cases with dashed dark red lines).  As in Fig.~\ref{fig:IndepBounds}, the dark photon gauge coupling is fixed to $\alpha_D=10^{-3}~(10^{-9})$ in the left (right) panel. 

Finally, in Fig.~\ref{fig:IndepBounds1} $\alpha_D$ is no longer fixed, but instead changes at each point to produce the correct relic abundance, Eq.~\eqref{eq:TD:alphaDeps}, assuming maximal dilution without internal thermalization, Eq.~\eqref{eq:Dmax:IT}.  The contours of $\alpha_D$ are denoted with diagonal dashed blue lines. The DM mass is  $m_\chi=1~(100)$ keV  in the left (right) panel, while $T_{\rm RH}= 2$~MeV. The blue shaded regions in the bottom right of the plot have $\alpha_D$ non-perturbative under our imposed assumptions. The left side of the figures is cut off by the requirement $T_D < m_{A'}/3$ in order to trust the behavior of the collision term.

\begin{figure}
  \centering
  \includegraphics[width=.49\linewidth]{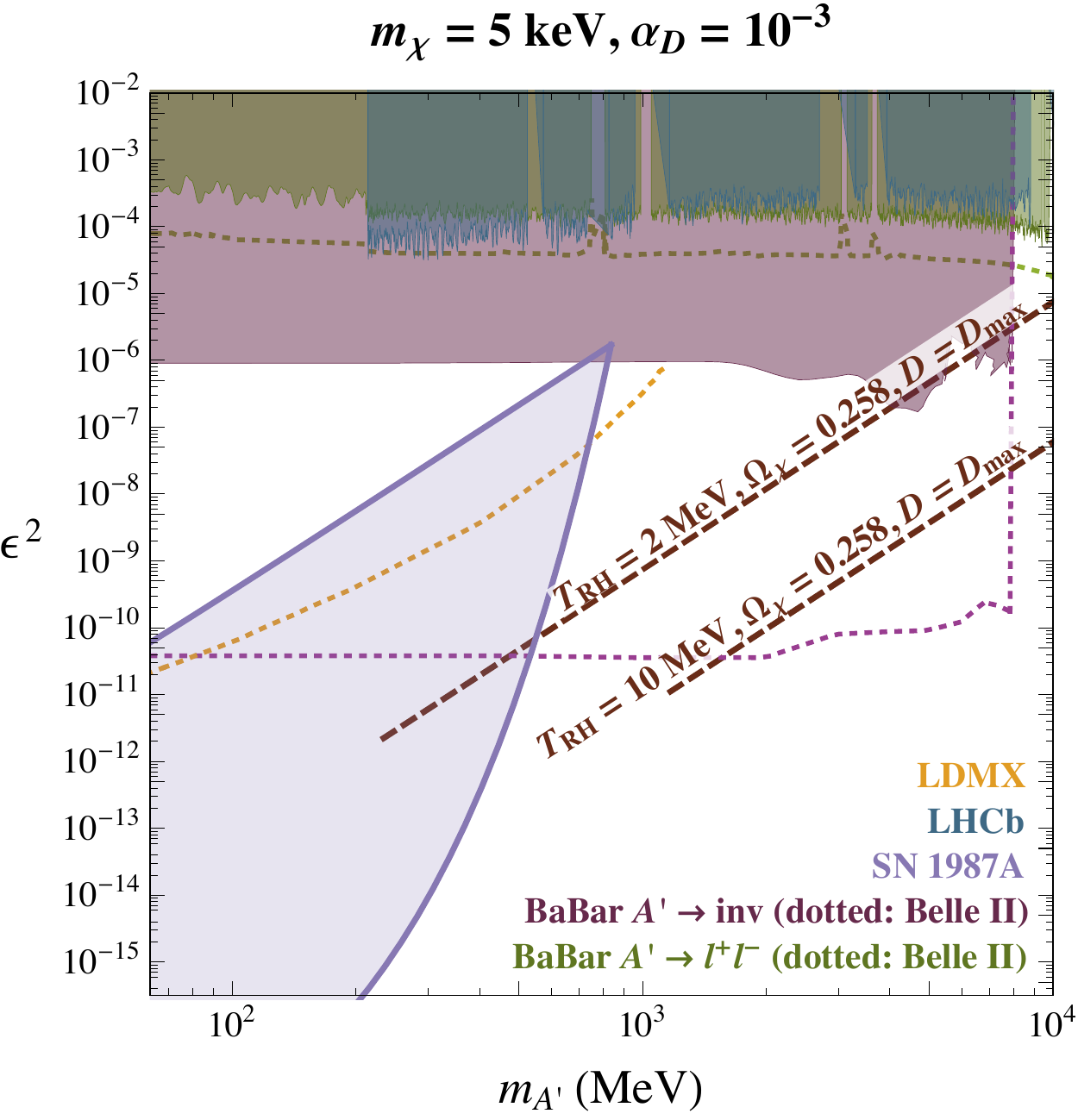}
  \includegraphics[width=.49\linewidth]{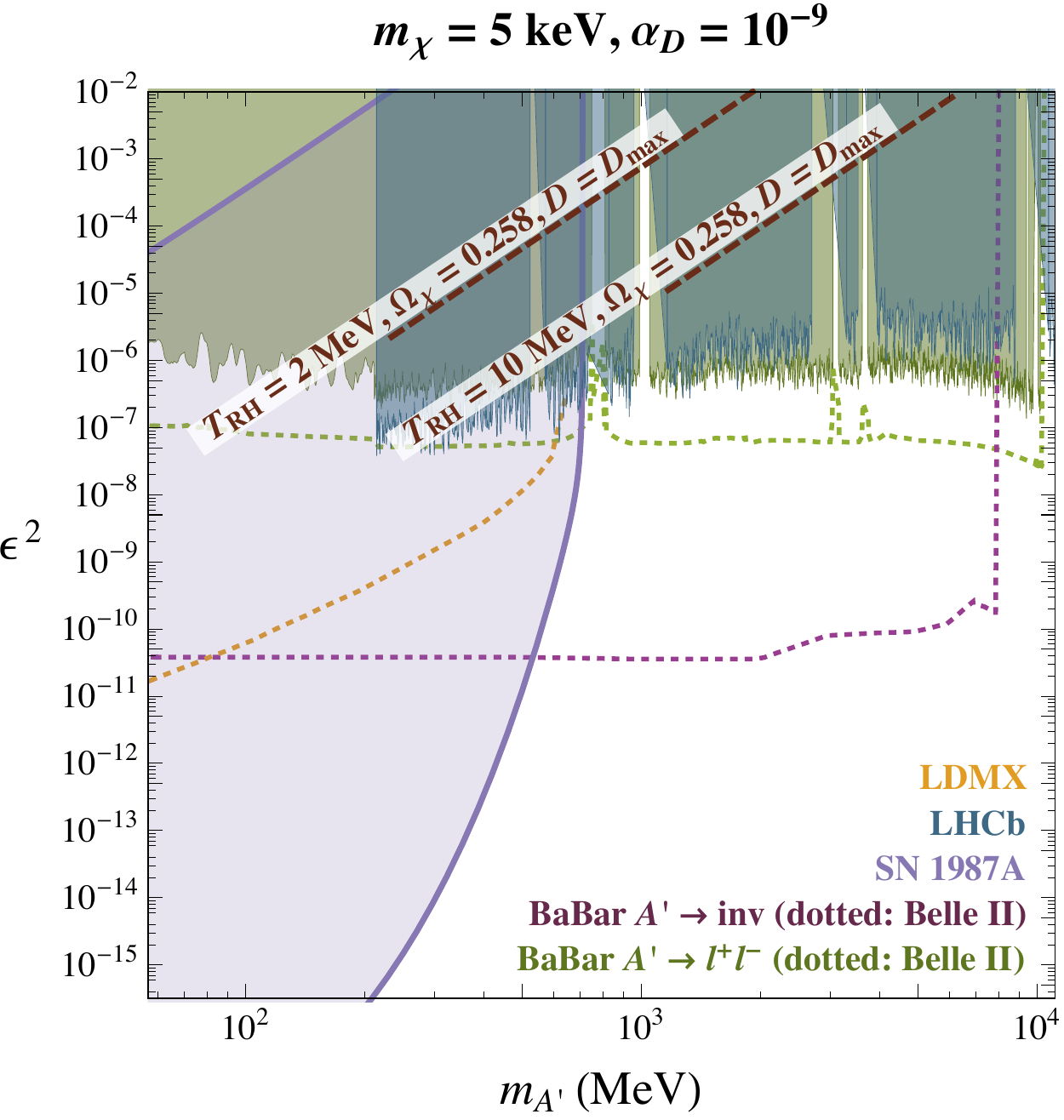}
  \caption{The same as in Fig.~\ref{fig:IndepBounds} except for the correct diluted hot relic DM abundance contours (dashed dark red lines) that are obtained for two different values of reheat temperature, $T_{\rm RH}=2$ MeV and $T_{\rm RH}=10$ MeV, by assuming maximal dilution with no internal thermalization \eqref{eq:Dmax:IT}.  
  }
\label{fig:IndepBoundsRH}
\end{figure}

The diluted hot relic DM scenario with a heavy vector portal can be probed in terrestrial experiments by searching for dark photons,  either in visible or invisible decay channels. The green shaded areas in Figs.~\ref{fig:IndepBounds}~--~\ref{fig:IndepBounds1} are probed
by the BaBar search for leptonic decays of a prompt dark photon in $e^+e^-\to\gamma A^\prime, A^\prime\to\ell^+\ell^-$, where $\ell=e,\mu$ \cite{Lees:2014xha}. For a dark photon that decays exclusively into SM states, this gives a bound on $\epsilon$ in the range $(5-10)\times 10^{-4}$. In the case of diluted hot relic DM,  the $A'\to \bar \chi \chi$ decays have also a sizable branching ratio, modifying, therefore, the reach on $\epsilon$. 
  As expected, in the case of $\alpha_D=10^{-9}$ (right panels in Figs. \ref{fig:IndepBounds} and \ref{fig:IndepBoundsRH}), the bound is quite close to the bound obtained with assuming 100$\%$ dark photon visible decays, since in that case $\alpha_{\rm{em}}\epsilon^2\gg\alpha_D$. 
   The exclusion from the  LHCb search for visibly decaying dark photons in the $A'\to\mu^+\mu^-$ \cite{LHCbtalk} channel is shown in blue, and is the most sensitive for light $A'$.  The parameter space of visibly decaying dark photons can also be constrained by fixed-target beam-dump experiments like  E137 \cite{Bjorken:1988as,Andreas:2012mt}, LSND \cite{Athanassopoulos:1997er,Essig:2010gu}, U70 \cite{Blumlein:2013cua}, CHARM \cite{Bergsma:1985qz,Gninenko:2012eq}, and SeaQuest \cite{Gardner:2015wea,Berlin:2018tvf,Berlin:2018pwi,Tsai:2019mtm}, if the dark photon is long lived enough to decay after the dump. However, we have checked that these experiments do not constrain additional regions of parameter space of our model, due to the relatively short life time of the dark photon in the region of interest.
    
    \begin{figure}
  \centering
  \includegraphics[width=.49\linewidth]{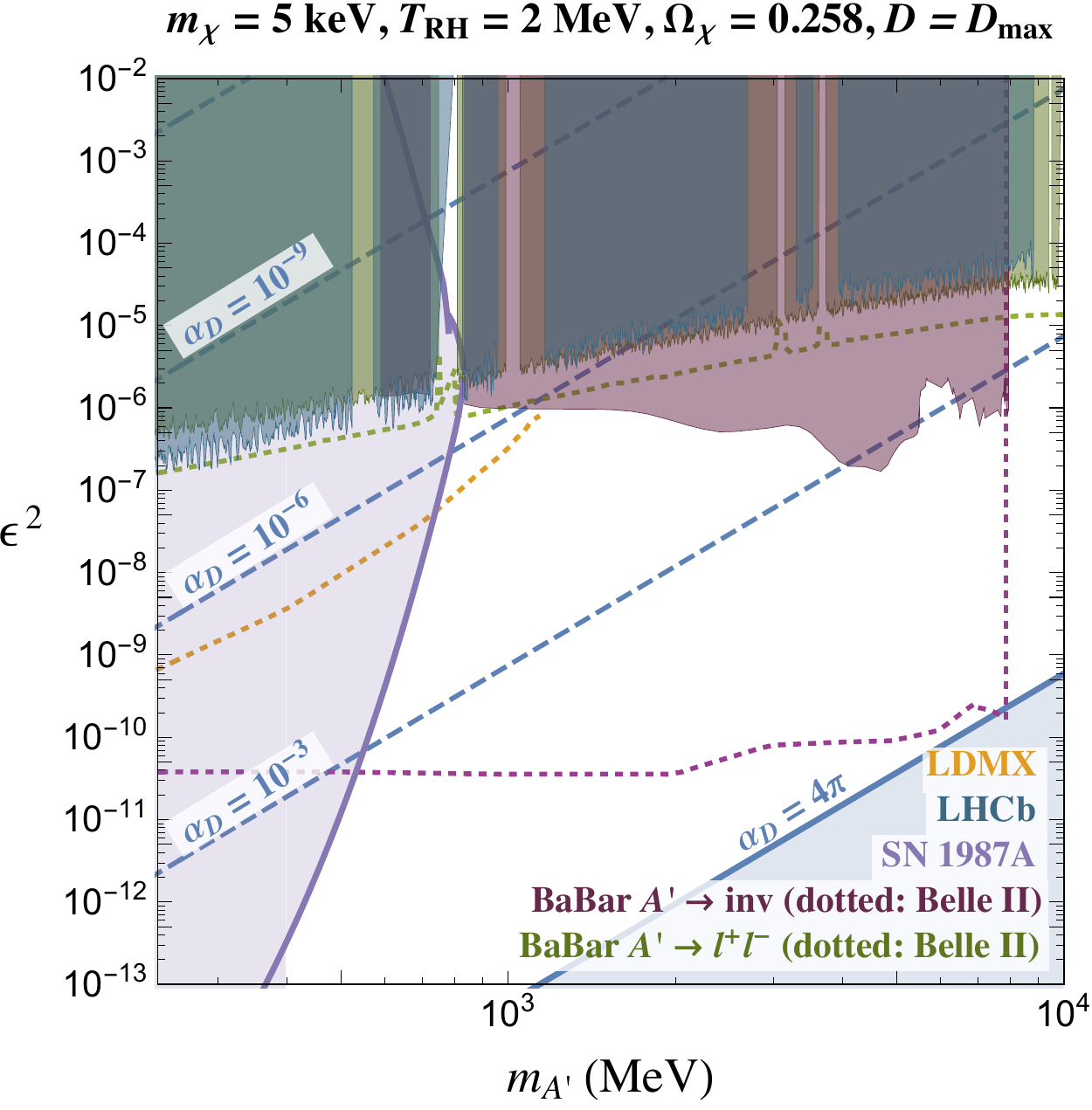}
  \includegraphics[width=.49\linewidth]{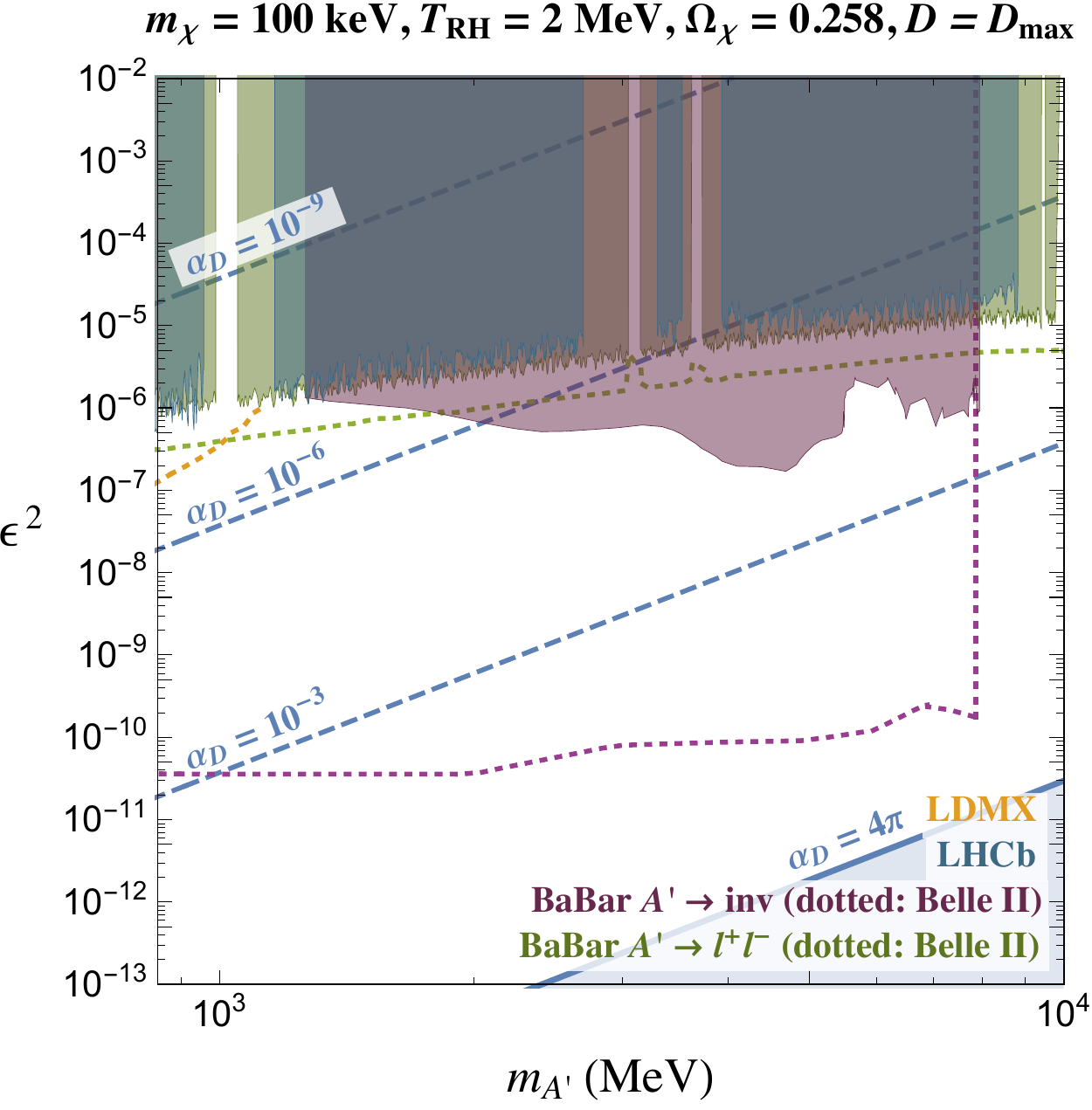}
\caption{The left (right) panel shows constraints on the diluted hot relic DM relic for $m_\chi=5\,(100)$ keV, taking the reheat temperature to be $T_{\rm RH}=2$~MeV. 
In contrast to Fig.~\ref{fig:IndepBounds} and Fig.~\ref{fig:IndepBoundsRH}, $\alpha_D$  changes at each point in the plane to produce the correct relic abundance assuming maximal dilution without internal thermalization.   
The contours of $\alpha_D$ are denoted by blue dashed lines, with the blue shaded region in the bottom right of each panel indicating where $\alpha_D$ is non-perturbative. 
The color coding for the experimental exclusions is the same as in Fig.~\ref{fig:IndepBounds}.
} 
\label{fig:IndepBounds1}
\end{figure}

BaBar also searched for invisibly decaying dark photons in the mono-photon channel $e^+e^-\to\gamma A^\prime, A^\prime\to$ inv \cite{Lees:2017lec}, leading to the maroon shaded exclusion regions in Figs.~\ref{fig:IndepBounds}~--~\ref{fig:IndepBounds1}. Assuming that the dark photon decays exclusively to invisible particles, the bound on $\epsilon$ is not too different from the one obtained from the visible decays.  The differences in bounds shown in Figs.~\ref{fig:IndepBounds}~--~\ref{fig:IndepBounds1} obtained from visible~vs.~invisible channels are thus entirely due to how large $\alpha_D$ is assumed to be. 

The projection for a bound on $\epsilon$ from monophoton searches at
 Belle II, utilizing 50 ab$^{-1}$ of data, taking into account the improved hermiticity of the Belle II detector compared to BaBar, is shown by the purple dotted lines in Figs.~\ref{fig:IndepBounds}~--~\ref{fig:IndepBounds1}
 ~\cite{Kou:2018nap,Dolan:2017osp}. Correspondingly, we also show the projected Belle II with 50 ab$^{-1}$ reach on the $e^+e^-\to\gamma A^\prime, A^\prime\to\ell^+\ell^-$ signature (dotted green line in the figures).  We note that in Figs.~\ref{fig:IndepBounds}~--~\ref{fig:IndepBounds1} the improvement in sensitivity of Belle II for invisibly decaying dark photon is larger than the visibly decaying one. Much of this difference comes from the fact that both for $\alpha_D=10^{-3}$ and $\alpha_D=10^{-9}$ Belle II will probe values of $\epsilon$ small enough that the dark photon almost exclusively decays to DM pairs, cf. Fig. \ref{fig:BRApAndLifeTime}. The sensitivity to visible decays is correspondingly reduced. The remaining difference in the reach is in part due to the Belle II detector being more hermetic than BaBar, and in part because the visible channel has an irreducible SM background.

For $A'$ with mass $m_{A'}\lesssim 20$ MeV, i.e., lighter than the range shown in Figs.~\ref{fig:IndepBounds}~--~\ref{fig:IndepBounds1}, the most stringent bounds on invisibly decaying dark photon come from the NA64 experiment~\cite{NA64:2019imj}. This is a fixed-target experiment at the CERN SPS searching for dark sector invisible signatures through the precision measurement of electrons scattering on a nucleus, 
 $e^- Z \to e^- Z A^\prime,~A^\prime\to$ invisible. Assuming that the $A^\prime$ decays  invisibly 100$\%$ of the time, the present bound on $\epsilon$ is as small as $\{$few$\}\times 10^{-6}$ for $m_{A^\prime}\sim$ MeV \cite{NA64:2019imj}.  However, NA64 is not sensitive enough to constrain any of our parameter space of interest.  Similarly, the proposed Light Dark Matter eXperiment (LDMX) employs missing momentum and energy techniques to search for invisible dark sector signatures. This experiment aims at extending the NA64 sensitivity by $\sim$ three orders of magnitude in coupling $\epsilon$ with $10^{16}$ electrons on target using an $8$ GeV beam \cite{Akesson:2018vlm}. The projected reach is shown with orange dotted lines in Figs.~\ref{fig:IndepBounds}~--~\ref{fig:IndepBounds1}. It is expected to significantly extend the reach for light  dark photons. 

The model also faces constraints from astrophysics. Since the benchmark DM masses are less than an MeV,  
the thermal production of DM particles through dark photon exchanges can cool stars (the Sun, red giants, and horizontal branch stars) or the proto-neutron star within a supernova. Except for the proto-neutron star within a supernova, the remaining stars have a temperature of at most $\sim 10 \, \keV$. Since our dark photon has mass $m_{A^{\prime}} \gtrsim 100 \,\MeV$, the production of DM in the stars with lower temperatures will be through an off-shell dark photon. We can compare this to the neutrino production in those stars 
\begin{align}
\frac{\Gamma_{\chi}}{\Gamma_{\nu}} \approx \frac{\epsilon^{2}\alpha_{\text{em}}\alpha_{D} m^{4}_{W}}{\alpha_W^2 m^{4}_{A^{\prime}}} \approx 3\times 10^{-4},
\end{align}
where $m_{W}$ is the mass of the $W$ weak boson and in the second equality we have noted that ${\epsilon^{2}\alpha_{D}}/{m^{4}_{A'}} \sim \TeV^{-4}$ for light DM (see Fig.~\ref{fig:IndepBounds1}).  The stellar cooling from dark matter emission is much lower than for neutrino emission, and therefore do not constrain our model. The proto-neutron stars on the other hand, can have temperatures comparable to the dark photon mass in our model. The cooling of a proto-neutron star thus has observational consequences; a large enough cooling shortens the duration of the neutrino pulse from the SN explosion.  The observed neutrino pulse from SN1987A can therefore be used to constrain the diluted DM model \cite{Chang:2016ntp,Chang:2018rso}. 

There are two relevant DM production mechanisms in a proto-neutron star. The first is the decay of the SM photon into DM due to its mixing with the dark photon (SM photon has plasma induced mass). The second production mechanism is the bremsstrahlung of DM pairs. For our model benchmark values we have $m_{\chi} \ll m_{A^\prime}$ and $m_{A^\prime} \gtrsim T_{c}$, where $T_{c}$ is the neutron star core temperature. In this regime the dominant DM emission mechanism is the 
bremsstrahlung of DM pairs. 

The resulting exclusion regions are shaded purple in Figs.~\ref{fig:IndepBounds}~--~\ref{fig:IndepBounds1}. The regions have upper and lower boundaries in $\epsilon$ and an upper boundary in $m_{A'}$.
For sufficiently low $\epsilon$ too few DM pair are produced to sufficiently modify the cooling of the proto-neutron star within SN1987A. For large enough $\epsilon$ DM interacts strongly enough with the medium that it does not escape the 
proto-neutron star. For large enough $m_{A'}$ the cooling mechanism shuts off. Further details on the calculation of the bounds
are given in Appendix \ref{app:SNE}. Note that these constraints rely on the assumptions regarding the mechanism underlying the supernova explosion and can thus be viewed as less reliable then the constraints from the terrestrial experiments. In particular, if the supernova explosion is not due to delayed neutrino mechanism, but rather due to the collapse-induced thermonuclear explosion, the SN1987A bound on free-streaming particles, such as the dark photon, is completely absent \cite{Bar:2019ifz}. 

From Figs.~\ref{fig:IndepBounds}~--~\ref{fig:IndepBounds1}, we can conclude that at small enough values of $\alpha_D$, our benchmark scenarios are extensively probed by terrestrial experiments. Larger values of $\alpha_D$ can also be probed by a combination of terrestrial and astrophysical measurements at large and small values of $m_{A^\prime}$, respectively.
Note that Figs.~\ref{fig:IndepBounds}~--~\ref{fig:IndepBounds1} show bounds for the benchmarks (\ref{eq:mass:benchmarks}), (\ref{eq:couplings:benchmarks}), which were chosen to maximize the signals in terrestrial experiments, hence the constraints are not entirely generic. 

A way to avoid terrestrial constraints while keeping the dilution factors unchanged
is to increase the reheat temperature above 2~MeV, and then to also raise appropriately the decoupling temperature by reducing $\epsilon^2 \alpha_D/m_{A'}^4$, cf.~Eq.~\eqref{eq:TD:alphaDeps}. Furthermore, in our benchmarks we assumed that the moduli decay time is exactly such that the maximal dilution is achieved, cf. Eq.~\eqref{eq:TNAmax}. If this assumption is relaxed, the effective coupling of the SM to the HS can be further reduced (i.e., $T_D$ can be raised if $T_{\rm NA}$ is also modified, keeping $D$ unchanged), weakening in such a way the bounds from terrestrial experiments.

\section{Conclusions}
\label{sec:conclusion}
A generic challenge with thermal DM that couples weakly with the SM is that it can quite easily result in an overclosed universe, i.e., that there is too much DM left over after the freeze-out period. This problem is especially pronounced for models with light DM, with masses below ${\mathcal O}(1\text{~GeV})$. These generically require couplings to the SM that are large enough to account for a sizable DM annihilation, resulting in possibly stringent bounds from 
terrestrial experiments or from astrophysical observations. 

A simple solution to this problem is that, during the cosmological evolution, the DM relic abundance gets diluted. If this happens, thermal relic DM with small couplings to the SM becomes viable, and thus also easily evades the experimental constraints. In this paper, we explored a particular realization of such a diluted hot relic DM where the dark sector entropy dilution is caused by a heavy state, ``moduli'', that decays (almost) exclusively into the SM sector. 
The energy injected into the SM by this decay will heat up the SM relative to the hidden sector (HS), as long as the two are no longer in kinetic equilibrium. 

With this very simple set-up, one may be tempted to 
conclude that the predictive power is lost  
as far as the properties of the HS and moduli are concerned. However, there are certain properties that the HS needs to satisfy. First of all, the effective couplings to the SM should be sufficiently weak to allow the HS to fall out of  kinetic equilibrium early enough. 
Second, the coupling of the HS to the SM should also not be too weak, so that the thermalization of the HS with the SM does occur in the early universe. These two requirements are naturally satisfied by the SM-HS interaction through a massive mediator, such as a dark photon. 
For temperatures above the mass of the mediator, the renormalizable interactions between the hidden and visible sector keep the two in thermal equilibrium. For temperatures below the mass of the mediator, the mediator can be integrated out and the interactions between the SM and the HS are through higher dimension operators.  The effective interactions between the hidden and SM sectors decouple quickly below the mass of the mediator with a power law dependence on the temperature. 

In this paper, we calculated in a model independent fashion the maximum entropy dilution a hidden sector can experience for a given coupling between the SM and the HS. After the SM gets heated by the moduli decay, the coupling between the SM and the HS plasmas leads to the `leak-in' of the energy from the SM into the diluted HS, a heating which restricts the degree to which the HS can be diluted. 

To explore the implications of this mechanism for terrestrial experiments,  we focused in the second part of the paper on a particular model where the mediator between the HS and SM is a heavy kinetically-mixed dark photon. We showed that, under the assumption of maximum dilution, the model is under the lamppost of current and future experiments. A large part of the parameter space is or will be probed by past/present (Babar, LHCb), and future (Belle II, LDMX) terrestrial experiments and could well be discovered in the near future. 
Searches for an invisibly decaying mediator are one of the most sensitive probes of diluted hot relic DM models, and would likely be one of the first signatures of this model to appear. 

We expect this to be a generic feature in many diluted hidden sectors with light particles.  
Such dark sectors can possess, in fact, sizable couplings within the dark sector as well as sizable terrestrially accessible couplings between the HS and SM. 
Fairly light dark matter ($m_\chi\sim$~few~keV) is possible in part because the extremely cold nature of the hidden sector helps to insulate it both from constraints on warm dark matter and on the number of relativistic species present during BBN ($N_{\rm eff}$). 
Unlike the undiluted case, large separations in scale between the mediator and dark matter are possible without having to greatly enhance the couplings to the SM.   
As a result, different detection opportunities could be relevant to test regions of parameter space not typically producing the measured relic abundance in thermal relic dark matter models. An example are searches for relativistic weakly coupled states produced in beams that subsequently scatter off of neutrino detection experiments that may probe diluted hot relic models. A comprehensive study of such possibilities is beyond the scope of the present work, but would be an interesting future research direction.

The calculations in this work are largely applicable to other dark sector scenarios beside the dark photon model studied here, for instance to Higgs or neutrino-mediated dark matter scenarios. 
While we focused on light dark matter, heavy dark matter that decouples from the SM through non-relativistic freeze-out could also be diluted to obtain the observed relic abundance.  In this scenario, leak in effects are unimportant, and there should be no limits on how much dilution can be applied. Alternatively, completely decoupled sectors can also have their matter density diluted. Entropy dilution is an interesting mechanism that may bring the dark sector under new lampposts, making its phenomenological implications worthy of further consideration.

\section*{Acknowledgments}
\label{sec:acknowledgments}

The authors would like to thank K.~Blum, J-H.~Chang, J.~Cornell, J.~Dror, D.~Iakubovskyi, S.~McDermott, and Y.~Wong for helpful conversations and clarifications.  We thank J.~Cornell for comments on the draft.   
We are especially grateful to Ayuki Kamada for pointing out an issue in our treatment of free-streaming in the first version of this paper.
JAE, SG, and AG would like to thank the Aspen Center for Physics under NSF grant PHY-1607611 where part of this work was completed. AG would like to thank GGI for hospitality during the completion of part of this work. JZ would like to thank ITP Warsaw, Poland, where part of this work was completed. SG would like to thank the Kavli Institute of Theoretical Physics for hospitality during the completion of parts of this work, and corresponding support from the National Science Foundation under Grant No.~NSF PHY-1748958. JAE, MT and JZ acknowledge support in part by the DOE grant DE-SC0011784. The research of SG and AG is supported in part by the NSF CAREER grant PHY-1915852. 

\begin{appendix}
\section{DM Production and Trapping in Supernovae}
\label{app:SNE}

In this Appendix, we discuss in detail DM  production in supernovae, and derive bounds on the vector portal model that follow from observations of supernova 1987A \cite{Chang:2016ntp,Chang:2018rso}. In a supernova explosion, there are two main DM production mechanisms: DM pair production via bremsstrahlung or DM production from decays of the SM photons in the thermal bath. The DM bremsstrahlung initiated by production of on-shell dark photons dominates over the SM photon induced production in our parameter space of interest. As the decays of SM photons to DM pairs lead only to small corrections, we will ignore their effects. Below, we review the procedure used to estimate the DM bremsstrahlung in a supernova, as well as the method used to estimate the trapping of DM inside a supernova, and apply it to the case at hand.

\subsection{DM pairs from dark photon bremsstrahlung}
Within the hot core of a supernova, DM pairs can be thermally produced, particularly, if the dark photon is sufficiently light to be accessible in collisions, $m_{A^\prime}\lesssim 500$ MeV.  The resulting DM bremsstrahlung luminosity, $L^{\rm brem}_{\chi} = \int^{R_{\nu}}_{0}dV{dL_{\chi}^{\rm brem}}/{dV}$, can be written as (see Appendix B in Ref. \cite{Chang:2018rso}) 
\beq
\begin{split}
\label{eq:brem luminosity}
L_{\chi }^{\text{brem}} =& \int_{0}^{R_{\nu}} dr 4\pi r^{2} \frac{\alpha_{em} \alpha_{D} \epsilon^{2}}{3\pi^{2}} \frac{16}{\sqrt{\pi}} n_{n}(r)n_{p}(r) \langle \sigma^{(2)}_{np} (T)\rangle \int d\cos(\theta_{k\chi}) dp_{\chi}dk \, p^{2}_{\chi}k^{2} \frac{\omega e^{\frac{-\omega}{T(r)}}}{\omega(\omega-p_{\chi})} \\
& \times \left\{\frac{\kappa^{4}}{\omega^{4}}\frac{\left(\kappa^{4}-4p^{2}_{\chi}\left(k-\omega \cos\theta_{k\chi}\right)^{2}\right)}{\left[\left(\kappa^{2}-m_{A'}^{2}\right)^{2}+\left(m_{A'}\Gamma_{A^\prime}\right)^{2}\right]\left[\left(\kappa^{2}-\text{Re}\,\Pi_{L}\right)^{2}+\text{Im}\,\Pi^{2}_{L}\right]} +\cdots \right\},
\end{split}
\eeq
where 
$\kappa = (\omega, \vec k)$ is the dark photon four-momentum, with $k=\big|\vec k\big|$, $\cos\theta_{k\chi}$ is the angle between the dark photon and dark matter three momenta,  while $R_{\nu} = 39.8$ km is the radius of the neutrinosphere.
We have set $m_{\chi} = 0$ in the above, since in our parameter space the DM mass is always smaller than the typical temperature in the supernova. 
For the temperature profile, $T(r)$, and the neutron (proton) number densities, $n_{n(p)}(r)$, 
 we use the fiducial profile functions in Eq.~(2.4) of Ref.~\cite{Chang:2016ntp}, including the numerical values for the parameters quoted there. 
In \eqref{eq:brem luminosity} we only kept the contributions from the longitudinal SM photon polarization, which in our case dominate the cross-section, with the ellipses denoting the sub-leading contribution from the transverse SM photon polarization and the cross terms. The function $\Pi_{L}$ gives the self-energy of the longitudinal SM photon\footnote{Note that alternative definitions for $\Pi_{L}$ are also used in the literature \cite{Braaten:1993jw}.} 
\begin{align}
\label{eq:pol}
\Pi_{L} & = 3\omega^{2}_{p}\left(\frac{\omega^{2}}{k^{2}}-1\right)\left(\frac{\omega}{2k}\log\frac{\omega+k}{\omega-k}-1\right),
\end{align}
where $\omega_{p}$ is the plasma frequency 
\beq
\omega^{2}_{p}= \frac{4\pi}{3}\left(\mu^{2}+\frac{\pi^{2}T^{2}}{3}\right),
\eeq
with $\mu$ the chemical potential of the electrons (equivalent to that of protons). The profile of the chemical potential, $\mu(r)$, follows from the assumed temperature profile, $T(r)$, and nucleon densities, $n_{n(p)}(r)$.

To evaluate \eqref{eq:brem luminosity}
we work in the narrow resonance width approximation, i.e., the dark photon is taken to be on-shell through the following  replacement,  $1/\big[(\kappa^2-m_{A'}^2)^2+(m_{A'}^2\Gamma_{A'})^2\big]\to \pi \delta(\kappa^2 -m_{A'}^2)/(m_{A'} \Gamma_{A'})$, in the integrand in \eqref{eq:brem luminosity}. The integration over $\cos{\theta_{k\chi}}$ then becomes trivial. We also assume $m_{A'}> 100\,\MeV$, so that $m_{A'} \gg \omega_{p}$ and we can safely neglect SM photon self energies, setting $\text{Re}\,\Pi_{T,L}  =\text{Im}\,\Pi_{T,L} = 0$. The $p_{\chi}$ integral can then be evaluated analytically.
 We perform instead the $k$ and $r$ integrations  numerically. For the thermally averaged neutron-proton dipole cross-section we take $\langle \sigma^{(2)}_{np} (T)\rangle = 100$\,mb irrespective of temperature \cite{Rrapaj:2015wgs}. Note that for $\alpha_{D}\gg \epsilon^{2}\alpha_{em}$ the DM luminosity $L_{\chi, {\rm brem}}$ does not depend on $\alpha_{D}$, since the $\alpha_D$ from the matrix element squared gets canceled by $1/\Gamma_{A^\prime}$. For $\alpha_{D}\ll \epsilon^{2}\alpha_{em}$, on the other hand $\Gamma_{A^\prime}$ is dominated by decays to visible sector, and thus $L_{\chi}^{ {\rm brem}}\propto{\alpha_{D}}/{\epsilon^{2}\alpha_{em}}$.
 
To derive an upper bound on $\epsilon^{2}$ for a given $\alpha_D$ we use the Raffelt criterion, requiring that the DM luminosity in \eqref{eq:brem luminosity} is less than the luminosity in neutrinos,  $L_{\chi}^{\rm brem} \leq L_{\nu} = 3\times 10^{52} {\rm ergs}/{\rm s}$.

\subsection{Trapping of DM inside supernova}

If the coupling between the SM and DM is large enough then DM remains trapped inside the supernova and does not contribute to the cooling of the proto-neutron star. This results in an upper bound on $\epsilon^{2}$. To calculate the scattering rate of DM inside the proto-neutron star we follow the method described in Ref. \cite{Chang:2018rso}. For this we use a simplified picture of the supernova; we assume that the DM is in thermal equilibrium inside the decoupling radius $R_d$ and free-streaming outside.  To determine $R_d$, we impose that the dark sector luminosity due to blackbody radiation at radius $R_d$, 
\begin{align}
L_{d} = L_{\chi}+L_{A^\prime} = 4\pi R^{2}_{d}\int dp \left(\frac{g_{\chi}}{8\pi^{2}}\frac{p^{3}}{e^{E_{\chi}/T}+1}+\frac{g_{A^\prime}}{8\pi^{2}}\frac{p^{3}}{e^{E_{A^\prime}/T}-1}\right),
\end{align}
equals the 
neutrino luminosity $L_\nu$. 
 In writing the expressions for the two pieces, we used that both $\chi$ and $A'$ are relativistic at $R_d$.
  The luminosities depend on the temperature profiles $T(r)$ for which we use the fiducial temperature profile in Eq.~(2.4) of Ref.~\cite{Chang:2016ntp}.

In order to find the value of $\epsilon$ above which DM gets trapped, we require that the  deflection angle for a typical DM particle departing the decoupling radius is $\langle|\theta\left(\alpha_{D},R_{d},\epsilon\right)|\rangle\geq\pi/2$, i.e., that it typically deflects completely. From the properties of a random walk in three dimensions we have \cite{Chang:2018rso} 
\begin{align}
\langle|\theta\left(\alpha_{D},R_{d},\epsilon\right)|\rangle = \frac{\theta_{\rm max}\left(\alpha_{D},R_{d},\epsilon\right)}{2}\sqrt{\frac{\pi}{N\left(\alpha_{D},R_{d},\epsilon\right)}} \geq \frac{\pi}{2},
\end{align}
where $N$ is the number of scatterings experienced by the particle on a trajectory from $R_d$ to the far radius $R_f$, and  $\theta_{\rm max}$ is the maximal angular deflection per each scattering. These quantities can be expressed as
\cite{Chang:2018rso}
\begin{align}
N\left(	\alpha_{D},\epsilon,R_{d}\right)  &= \int^{R_{f}}_{R_{d}}\frac{dr \Gamma_{s}\left(\alpha_{D},\epsilon,\bar{E}(R_{d}),r\right)}{v_\chi},  \\ 
\theta_{\rm max} \left(	\alpha_{D},\epsilon,R_{d}\right)  &= \int^{R_{f}}_{R_{d}}\frac{dr \Gamma_{s}\left(\alpha_{D},\epsilon,\bar{E}(R_{d}),r\right)\Delta\theta}{v_\chi}.
\end{align}
 Since DM is relativistic we can set the average DM velocity to $v_\chi=1$.  The thermally averaged energy at the decoupling radius  $\bar E(R_d)$ gives the initial DM energy. To determine its value we use the fiducial temperature profile in Eq.~(2.4) of Ref.~\cite{Chang:2016ntp}.
For the ``far radius," beyond which neutrinos are not effectively produced, we take $R_{f} = 100 \,\rm km$.

The $\chi + p \rightarrow \chi + p$ scattering rate,  $\Gamma_{s}$, and the average angular deflection per scattering, $\Delta\theta$,
are given by
\begin{align}
\label{eq:rate}
\Gamma_{s} &= \frac{1}{2E_{1}}\int \frac{d^{3}p_{2}}{2E_{2}}\frac{d^{3}p_{3}}{2E_{3}}\frac{d^{3}p_{4}}{2E_{4}}(2\pi)^{4}\delta^{4}\left(P_{1}+P_{2}-P_{3}-P_{4}\right) f_{2} |M_s|^{2},\\
\Delta \theta &= \frac{1}{2E_{1}\Gamma_{s}}\int \frac{d^{3}p_{2}}{2E_{2}}\frac{d^{3}p_{3}}{2E_{3}}\frac{d^{3}p_{4}}{2E_{4}}(2\pi)^{4}\delta^{4}\left(P_{1}+P_{2}-P_{3}-P_{4}\right)\theta_{13} f_{2} |M_s|^{2},
\end{align}
where $f_{2} = n_{p}\left({2\pi}/{m_{N}T}\right)^{3/2}e^{-p^{2}/2 m_{N}T}$ 
is the Maxwell-Boltzmann distribution for protons. The amplitude squared for the $\chi (P_1) + p(P_2) \rightarrow \chi(P_3) + p(P_4)$ scattering is given by \cite{Chang:2018rso}
\beq
\begin{split}
\label{eq:scatamp}
|M_s|^{2} &= \frac{16 \pi^{2}\epsilon^{2}\alpha \alpha_{D} K^{4}}{\left(K^{2}-m_{A'}^{2}\right)^{2}+\left(m_{A'}\Gamma_{\chi}\right)^{2}}\left(\frac{P_{L\mu\nu}}{K^{2}-\Pi_{L}}+\frac{P_{T\mu\nu}}{K^{2}-\Pi_{T}}\right)\left(\frac{P_{L\alpha\beta}}{K^{2}-\Pi^{*}_{L}}+\frac{P_{T\alpha\beta}}{K^{2}-\Pi^{*}_{T}}\right)\\
&\quad\times \Tr\left[\gamma^{\mu}\left(\slashed{P}_{1}+m_{\chi}\right)\gamma^{\alpha}\left(\slashed{P}_{3}+m_{\chi}\right)\right]\Tr\left[\gamma^{\nu}\left(\slashed{P}_{2}+m_{N}\right)\gamma^{\beta}\left(\slashed{P}_{4}+m_{N}\right)\right],
\end{split}
\eeq
where $\Pi_{L,T}$ are the self-energies for the SM longitudinal and transverse polarization. The transverse and longitudinal projection operators for the dark photon, $P_{T,L}^{\mu\nu}$, are given by
\begin{align}
\label{eq:projection}
P_{T\mu\nu} &= (1-\delta_{\mu,0})(1-\delta_{\nu,0})(\delta_{i,j}-k_{i}k_{j}/\vec{k}\cdot\vec{k}),\\
P_{L\mu\nu} &= -g_{\mu\nu}+K_{\mu}K_{\nu}/K\cdot K + P^{T}_{\mu\nu}.
\end{align}
We find that the longitudinal part of the amplitude gives the largest contribution to the scattering rate.

\section{Internal Thermalization of the Dark Sector}
\label{app:IT}
In the main text, we presented a derivation of the maximum allowed dilution, $\bar D_{\rm max}$, for the case where internal thermalization of the dark sector is maintained until the reheat temperature, $T_{\rm RH}$, see Eq.~\eqref{eq:Dmax}.  When internal thermalization is maintained, energy injected from the SM into the HS plasma rapidly equilibrates.  Conceptually, this means that a single DM particle injected with energy $T$ quickly converts into \emph{several} DM particles of energy $\tilde T$, with the number of particles given by $\sim T/\tilde T$ (we are interested in the case where the DM particles are relativistic during decoupling).  In contrast, if internal thermalization is absent, then there is no well-defined HS plasma. 
Below the temperature $T_{\rm IT}$ at which the internal thermalization ceases, the injected high energy HS particles do not get converted into many particles and their excess energy simply redshifts away.  
Without internal thermalization of the HS the overall number density of injected DM particles is therefore lower than when the sector is internally thermalized, and thus a smaller amount of dilution is required to obtain the correct DM relic abundance.   In this appendix, we discuss in detail the requirements for the HS to maintain internal thermal equilibrium. We also derive an expression for the maximum dilution, $D_{\rm max}$, that is valid when the HS is not in internal thermal equilibrium, cf.~Eq.~\eqref{eq:Dmax:IT}.  
 Over most of the parameter space of our benchmark model, internal thermalization is not maintained, and the procedure derived here is the one used to determine the maximum dilution throughout the main text.

We start by quantifying how large $\alpha_{D}$ needs to be in order for the HS to maintain internal thermal equilibrium throughout the relevant cosmological evolution.  Assuming that the HS thermalization occurs predominantly through a higher dimension operator ${\mathcal O}$ of  dimension $(n/2+4)$,  such that the effective interaction Lagrangian is ${\cal L}_{\rm eff}= \alpha_D {\mathcal O}/M^{n/2}$, the HS thermalization rate is parametrically given by
\beq
\label{eq:dtherm}
\Gamma_{\rm HS}(\tilde T) \sim n_{\chi}(\tilde T)\frac{\pi \alpha^{2}_{D}\tilde T^{n-2}}{M^n}.
\eeq
  In the case of the massive dark photon, 
   $n=4$, while $\alpha_D$ is the dark sector fine structure constant, and $M=m_{A'}$.  
   
    In order for the expression for maximum dilution in Eq.~\eqref{eq:Dmax} to be valid, the thermalization rate $\Gamma_{\rm HS}(\tilde T) $ needs to be larger than the Hubble expansion rate at $T_{\rm RH}$ (and thus also at all higher temperatures),
\beq
\label{eq:HTRH}
H(T_{\rm RH}) \simeq \frac{1}{M_{p}}\sqrt{\frac{4\pi^{3}g_*(T_{\rm RH} )}{45}}T_{\rm RH}^{2}.
\eeq
For simplicity, we neglect the  HS contributions to the Hubble expansion rate, since these are much smaller.
Equating \eqref{eq:dtherm} and \eqref{eq:HTRH} gives 
\beq
\label{eq:alphaDeq}
\alpha^{2}_{D,{\rm eq}}(a_{\rm RH}) =\frac{1}{\zeta_3 g_\chi}\sqrt{\frac{4\pi^5 g_*(T_{\rm RH} )}{45}}\frac{T_{\rm RH}^2}{\tilde T_{\rm RH}^{n+1}}\frac{M^n}{M_{p}}.
\eeq
 If $\alpha_D > \alpha_{D,{\rm eq}}(a_{\rm RH})$, then the HS maintains internal thermal equilibrium throughout the cosmological evolution up to and including the reheat time $t_{\rm RH}$, i.e., until the scale parameter reaches $a_{\rm RH}$, and therefore the calculation leading to Eq.~\eqref{eq:Dmax} is  consistent. Similarly, we can define $\alpha_{D,{\rm eq}}(a)$ for any other moment, by requiring that $\Gamma_{\rm HS}(\tilde T)|_a=H(T)|_a$. For $\alpha_D=\alpha_{D,{\rm eq}}(a)$ the HS is in internal equilibrium until the moment when the scale parameter reaches the value $a$.
 
It is instructive to express $\alpha^{2}_{D,{\rm eq}}(a_{\rm RH})$ in terms of our input parameters and estimate its typical numerical size. Using 
\eqref{eq:Ddef} and \eqref{eq:mvsD},  we can rewrite the above expression for $\alpha_{D,{\rm eq}}$ as
\beq
\label{eq:alphaDeq:2}
\alpha^{2}_{D,{\rm eq}}(a_{\rm RH}) =\frac{1}{\zeta_3}\sqrt{\frac{4\pi^5 }{45}}\frac{g^{1/2}_*(T_{\rm RH} )}{ g_\chi} \lp\frac{\tilde g_*(\tilde T_{\rm RH} )}{g_*(  T_{\rm RH} )} \rp^{\frac{n+1}{3}} \frac{ r^n T_D^n}{T^{n-1}_{\rm RH}M_{p}} \lp \frac{m_\chi}{\eta\, 1.5\mbox{ eV}}\rp^{\frac{n+1}{3}},
\eeq
where we defined
\beq
r\equiv \frac{M}{T_D},
\eeq
which is a measure of how far below the relevant mass scale decoupling occurs.  
In the case of our heavy vector model, we expect $r\gtrsim 3$.
If the dilution reaches its maximal value \eqref{eq:Dmax} we have ${T_{\rm RH}}/{T_{D}}=\left({\bar \lambda_D}/{\bar D_{\rm max}}\right)^{1/{\bar \gamma_n}}$.  We then have
\beq
\alpha_{D,{\rm eq}}(a_{\rm RH})= \frac{2.08}{ g_\chi^{1/2}} r^{n/2} \bar \lambda_D^{-\frac{n}{2 \bar \gamma_n}} \lp \frac{\tilde g_*^{n+1}(\tilde T_{\rm RH} )}{ g_*^{(2n-1)/2}( T_{\rm RH} )} \rp^{1/6}
\! \lp\frac{T_{\rm RH}}{M_p}\rp^{1/2}
\! \lp \frac{m_\chi}{\eta\, 1.5\mbox{ eV}}\rp^{\frac{n}{2\bar \gamma_n} +\frac{(n+1)}{6}}\,.
\label{eq:alphaITcompare}
\eeq
As a useful numerical example let us take  
$n=4$,  $T_{\rm RH}=2$ MeV, and assume that DM is a Dirac fermion  so that $g_\chi=4$ and $\tilde g_{*}(\tilde T)=3.5$ over the range of interest.  The minimum value of $\alpha_D$ that will maintain internal thermal equilibrium is then
\beq
\alpha_{D,{\rm eq}}(a_{\rm RH}) =1.3 \times 10^{-4} \lp\frac{r}{3}\rp^2 \lp\frac{m_\chi}{\text{keV}}\rp^{\frac{79}{42}}.
\eeq
For $r=3$ and $m_\chi =5$ keV, then $ \alpha_{D,{\rm eq}}(a_{\rm RH}) \approx 2.7\times 10^{-3}$, while  if $m_\chi =100$ keV, then $ \alpha_{D,{\rm eq}}(a_{RH}) \approx 0.7$. The calculation performed in the main text is consistent for $\alpha_{D}$ above these values.

For couplings smaller  than $\alpha_{D,{\rm eq}}(a_{\rm RH})$, the HS is not in internal thermal equilibrium throughout the relevant cosmological evolution.  
Without internal thermalization, the useful quantity to calculate is the non-thermal DM \emph{number} density injected into the HS sector as opposed to the energy density.   
The excess energy density injected will eventually redshift away.  To this end, let us consider the extremal case  where the HS falls out of thermal equilibrium with itself \emph{before} the energy injected into the HS exceeds the red-shifting energy, i.e., at a temperature $T_{\rm bal}$, where $T_{\rm bal} < T_{\rm NA}$ is defined by
\beq
4 H(T_{\rm bal}) \tilde \rho(\tilde T_{\rm bal}) = C_E( T_{\rm bal}),
\eeq
so that the two terms governing the relation in \eqref{eq:maxratio} are once again balanced.\footnote{There is no well-defined HS plasma temperature without internal thermal equilibrium, so our use of HS temperatures below $\tilde T_{\rm IT}$ may seem cause for concern.   However, until $T_{\rm bal}$, the HS is simply redshifting, so that $\tilde \rho \propto a^{-4}$, which is exactly how a na\"ively defined HS temperature would scale.  For this reason, we can continue to treat the system as if it has a temperature.}  
Once this condition is imposed, the adiabatic and non-adiabatic evolutions of the two sectors from $T_D$ can be used to derive (for $T_D>T_{\rm NA}$)
\beqa
T_{\rm bal} &=& \lp\frac{\tilde g_{*}(T_D)}{\tilde g_{*}(\tilde T_{\rm bal})}\rp^{\frac1{29-3n}}  \lp\frac{g_{*}^2(T_{\rm bal})}{g_{*}(T_{\rm NA})g_{*}(T_D)}\rp^{\frac{11}{2(29-3n)}} T_{\rm NA}^{\frac{55}{2(29-3n)}}T_D^{\frac{3-6n}{2(29-3n)}} \\
&\overset{n\to 4}{=} & \lp\frac{\tilde g_{*}(T_D)}{\tilde g_{*}(\tilde T_{\rm bal})}\rp^{\frac1{17}}  \lp\frac{g_{*}^2(T_{\rm bal})}{g_{*}(T_{\rm NA})g_{*}(T_D)}\rp^{\frac{11}{34}} T_{\rm NA}^{\frac{55}{34}}T_D^{-\frac{21}{34}}.
\eeqa
By requiring $\Gamma_{\rm HS}(\tilde T_{\rm bal})=H(T_{\rm bal})$, we can derive
\beqa
\alpha^{2}_{D,{\rm eq}}(a_{\rm bal}) &=&\sqrt{\frac{4\pi^5}{45\zeta_3^2}} \frac{M^n T_{\rm bal}^{\frac{4}{3}\lp 1-2n\rp}T_{\rm NA}^{\frac{5}{3}\lp n+1\rp}}{g_\chi M_{p}  T_{\rm RH}^2} \left[ \frac{\tilde g_{*}(T_D)}{\tilde g_{*}(T_{\rm bal})} \frac{g_{*}^2(T_{\rm bal})}{g_{*}(T_{\rm NA})g_{*}(T_D)} \right]^{-\lp n+1\rp/3}.
\eeqa
The scale $T_{\rm bal}$ and the subsequent evolution are illustrated in Fig.~\ref{fig:schematicplotApp}.

\begin{figure}[t]
\centering
 \includegraphics[scale=1.1]{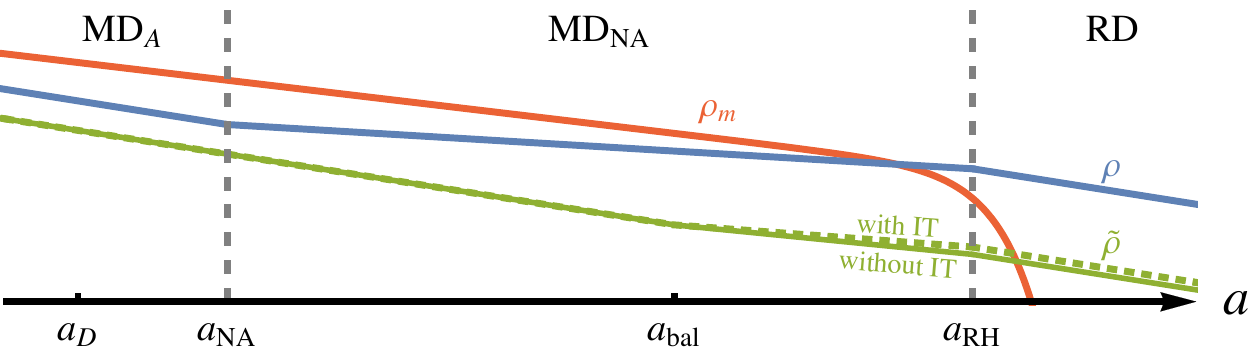}
\caption{The evolutions of the energy densities of the moduli, $\rho_m$ (red), the SM sector, $\rho$ (blue), and hidden sector, $\tilde \rho$ (green).  The $a_{i}$ are the scale factors corresponding to the temperatures $T_i$ and $\tilde T_i$.  Below the scale $a_{\rm bal}$, at which the red-shifted initial HS energy density becomes equal to the one injected from the SM, the HS energy density evolves slightly differently if the sector is or is not internally thermalized throughout the duration of the evolution up to $T_{\rm RH}$, i.e., $T_{\rm IT}< T_{\rm RH}$ or $T_{\rm IT}> T_{\rm bal}$, respectively.  The two cases are represented by dashed and solid lines, respectively.  All the other features in the diagram are as in Fig.~\ref{fig:schematicplot}.
  \label{fig:schematicplotApp}}
\end{figure}

Finally, we derive the expression for the maximum allowed dilution for $\alpha_{D}\leq \alpha_{D,{\rm eq}}(a_{\rm bal})$, so that the HS loses internal thermalization before $T_{\rm bal}$.  Without internal thermalization, the injected non-thermal DM number density can start to dominate over the thermal DM population which had been red-shifting ever since it had been in thermal equilibrium. From the expression for DM number density evolution 
\beq
\frac{d n_\chi}{dt} = -3 H n_\chi +C(T, \tilde T),
\eeq  
 we see that we can use $3 H n_\chi \geq C(T, \tilde T)$ as the criterion for when the non-thermal DM dominates, in close equivalence to the derivations in Section \ref{sec:HSdil}.  Here, $n_\chi$ initially comes from a redshifted thermal distribution, but below $T_{\rm bal}$, the injected DM particles do not track a thermal distribution.  Still, as long as $\tilde \rho /\tilde g_* \ll \rho/g_*$, the form of the phase space density will only influence the blocking or stimulated emission factors in the Boltzmann equation.  

Expressing the collision term as $C(T, \tilde T) \approx \kappa c_E T^{4+n} M^{-n}$ and using (\ref{eq:decouplingcond}), we then obtain
\beq
n_\chi (T) \geq \frac{C(T, \tilde T)}{3H(T)} =  \kappa \frac{2\pi^{2}}{45} \frac{H(T_D)}{H(T)} \tilde g_*(T_D)\lp\frac{T}{T_D}\rp^{n+1} T^3,
\label{eq:nchiIT}
\eeq
where $\kappa \equiv T C(T, \tilde T)/C_{E}(T, \tilde T) \sim \mathcal{O}(1)$.
As before,
\beq
\frac{H(T_D)}{H(T_{\rm RH})} = \frac{g_*^{\frac 12}(T_D)g_*^{\frac 12}(T_{\rm NA})}{g_*(T_{\rm RH})} \frac{T_D^{\frac 32}T_{\rm NA}^{\frac 52}}{T_{RH}^4},
\label{eq:Hratio}
\eeq
so from \eqref{eq:Ddef}, the maximum dilution becomes
\beq
D_{\rm max}= \frac{45 \zeta_3}{2\pi^4 \kappa \eta} \frac{g_*^2(T_{\rm RH})}{g_*^{\frac 12}(T_D)g_*^{\frac 12}(T_{\rm NA})\tilde g_*(T_D) }  \frac{T_{\rm RH}^{3-n}T_D^{n-\frac 12}}{ T_{\rm NA}^{\frac 52}},
\label{eq:maxdilIT1}
\eeq
where  $\eta = 7/6\, (1)$ for fermions (bosons).  To define the maximum dilution, we used the entropy of the HS, which may seem problematic without thermal equilibrium. However, the introduction of entropy is primarily used to track the redshifting of the DM density and hence will still reliably track the redshift and produce accurate results as long as the HS co-moving number density is conserved.
Eq.~\eqref{eq:maxdilIT1} is a weaker condition than the equilibrated case \eqref{eq:maxdil}.  As before, we can equate \eqref{eq:dil} and \eqref{eq:maxdilIT1}, and solve for $T_{\rm NA}$ to simplify the expression to give $D_{\rm max}$ in Eq.~\eqref{eq:Dmax:IT}.  This is the maximal possible dilution in the case when internal thermalization is lost before the particle injection become more important than the red-shift, i.e., $\tilde T_{\rm IT} > \tilde T_{\rm bal}$.  Note that $T_{\rm IT}\leq T_D$ always holds, as internal thermalization of the sector will be maintained while the SM and HS are in thermal equilibrium.\footnote{Non-thermal corrections may be important if  $ T_{\rm IT} = T_D$, see e.g. \cite{DAgnolo:2017dbv}. We can neglect this effect to the precision we are working.}

When $\alpha_{D,{\rm eq}}(a_{\rm RH}) <\alpha_{D}<\alpha_{D,{\rm eq}}(a_{\rm bal})$, the internal HS thermalization scale $T_{\rm IT}$ would enter in the expressions.  In this case, the maximum allowed dilution would sit between the completely internally thermalized and decoupled cases.  While the derivation of this term is straightforward, it is not very illuminating.

\end{appendix}

\bibliography{DDM}
\bibliographystyle{JHEP}
	
\end{document}